\documentclass[preprint,showpacs,preprintnumbers]{revtex4}
\usepackage{amsmath}
\usepackage{graphicx}
\usepackage{dcolumn}
\usepackage{bm}
\usepackage{epsfig}
\usepackage[T1]{fontenc}
\usepackage{ae,aecompl}
\usepackage{epstopdf, epsfig}
\usepackage{color}
\usepackage{appendix}

\begin{document}

\title{Discrete Boltzmann trans-scale modeling of high-speed compressible
flows}
\author{Yanbiao Gan$^{1,2}$, Aiguo Xu$^{3,4}$\thanks{
Corresponding author. E-mail: Xu\_Aiguo@iapcm.ac.cn}, Guangcai Zhang$^3$,
Yudong Zhang$^{3,5}$, Sauro Succi$^{6,7}$}
\affiliation{1, North China Institute of Aerospace Engineering, Langfang 065000, China\\
2, College of Mathematics and Informatics $\&$ FJKLMAA, Fujian Normal
University, Fuzhou 350007, China \\
3, National Key Laboratory of Computational Physics, Institute of Applied
Physics and Computational Mathematics, P. O. Box 8009-26, Beijing 100088,
China \\
4, Center for Applied Physics and Technology, MOE Key Center for High Energy
Density Physics Simulations, College of Engineering, Peking University,
Beijing 100871, China\\
5, Key Laboratory of Transient Physics, Nanjing University of Science and
Technology, Nanjing 210094, China\\
6, Center for Life Nano Science at La Sapienza, Fondazione Istituto Italiano
di Tecnologia, Viale Regina Margherita 295, 00161, Roma, Italy\\
7, Physics Department and Institute for Applied Computational Science, John
A. Paulson School of Applied Science and Engineering, Harvard University,
Oxford Street 29, Cambridge, MA 02138, USA }
\date{\today }

\begin{abstract}
We present a general framework for constructing trans-scale \emph{discrete
Boltzmann models} (DBMs) for high-speed compressible flows ranging from
continuum to transition regime. This is achieved by designing a higher-order
discrete equilibrium distribution function which satisfies additional
nonhydrodynamic kinetic moments. In order to characterize the \emph{%
thermodynamic non-equilibrium} (TNE) effects and estimate the condition
under which the DBMs at various levels should be used, two novel measures
are presented: (i) the relative TNE strength, describing the relative
strength of the ($N+1$)-th order TNE effects to the $N$-th order one; (ii)
the TNE discrepancy between DBM simulation and relevant theoretical
analysis. Whether or not the higher-order TNE effects should be taken into
account in the modeling and which level of DBM should be adopted, is best
described by the relative TNE intensity and/or the discrepancy, rather than
by the value of the Knudsen number. As a model example, a two-dimensional
DBM with $26$ discrete velocities at Burnett level is formulated, verified,
and validated.
\end{abstract}

\preprint{}
\maketitle

\section{Introduction}

High-speed compressible flows with substantial \emph{hydrodynamic and
thermodynamic non-equilibrium } (HNE and TNE, respectively) effects are
ubiquitous in nature, high pressure science and technology \cite{Xu-SciChina}%
, turbulent combustion \cite{Ju}, shock wave therapy \cite%
{Sterilization,medical-treatment}, food processing \cite{food}, hypersonic
flows associated with spacecraft reentry into the upper atmosphere \cite%
{2004-Li-JCP,2009-Li-JCP,2015-Li-PAS,2015-Wang}, and strong shock waves in
the inertial confinement fusion process \cite{He-SciChina,WLF}, etc.
More specifically, in the last two representative fields, both rarefied gas
flows at high altitude and limited shock wave thickness (typically of the
order of a few mean-free-paths of molecules, characterized by drastic
changes in state variables) give rise to high Knudsen number and significant
out-of-equilibrium scenarios. Additionally, most hypersonic vehicles operate
over a wide range of Knudsen number in different parts of the equipment,
simultaneously \cite{2004-JFM,2004-Li-JCP,2009-Li-JCP,2015-Li-PAS,2015-Wang}%
. Consequently, various flow regimes with totally different
aerothermodynamics, ranging from continuum, slip, transition, even to free
molecular flow regimes coexist in the entire flow, which adds considerably
to the complexity of the problem. For such complex non-equilibrium systems,
the appropriateness of constitutive relations, which are associated with the
TNE effects, ultimately determines the accuracy of the hydrodynamic model.
Besides the HNE effects described by hydrodynamic model, the evolution of
TNE characteristics helps to dynamically characterize the non-equilibrium
state and understand the constitutive relations. Therefore, establishing a
physically accurate and computationally efficient predictive model to
investigate these extremely complex HNE and TNE behaviors, is of both great
academic significance and industrial practical value. Undoubtedly, it is a
long-standing challenge.

The difficulty arises from the fact that various temporal and spatial scales
are associated and coupled with distinct physics. Consequently, the flow
lacks scale separation and the complexity springs up \cite{Succi-Science}.
Continuum-based Navier-Stokes (NS) equations, even with slip boundary
conditions, are not adequate to describe these kinds of flows. The
inadequacy stems from the linear constitutive relations for viscous stress
and heat flux assumed in the NS model which are no longer valid for the
far-from-equilibrium system. Thus, it is reasonable to conjecture that the
inclusion of higher-order terms in the constitutive relations can improve
the multi-scale predictive capability of such continuum models. Along this
line, the Burnett-like equations, which are expected to perform well in the
continuum-transition regime, are obtained from the CE expansion
of Boltzmann equation. Nevertheless, the extended hydrodynamic models are
still subject to at least the following four constraints that greatly hamper
their wide applications: (i) small wavelength instability as the grids are
refined; (ii) necessity of additional boundary conditions, (iii) complicated
programming owing to the existence of extraordinarily complex and numerous
higher-order derivatives, and (iv) the evolution equations of relevant
higher-order nonconservative kinetic moments are not included, even though
they are needed for an exact characterization of the non-equilibrium
behaviors and understanding the kinetic mechanisms for the nonlinear
constitutive relations. Currently, the particle-based direct simulation
Monte Carlo (DSMC) method has been regarded as a reliable and accurate
approach for simulating rarefied gas flows with high-speed and high Knudsen
number \cite{Caflisch1995,Caflisch1998,Caflisch1999}. Unfortunately, it
becomes extremely time-consuming and prohibitively memory-demanding for
simulating nearly continuum flows where intensive particle collisions take
place due to the limitation to the cell size and time step which are,
respectively, less than the mean-free-path and particle collision time. So,
it still cannot be qualified as a computationally efficient candidate for
modeling flows in the continuum-transition regime.

Being one of the most fundamental equations of the non-equilibrium
statistical physics, Boltzmann equation is capable of handing
thermohydrodynamics for the full spectrum of flow regimes. However, the
direct solution of the full Boltzmann equation encounters serious
difficulties due to the inherent nonlinearity, multidimensionality, together
with the multiple integro-differential nature of the collision term.
Therefore, developing approximate and simplified kinetic models which can
preserve the most relevant features of Boltzmann equation is currently an
important and essential attempt \cite%
{2004-Li-JCP,2009-Li-JCP,2015-Li-PAS,Succi-Book,Succi-2038,Struchtrup,XuKun,Guo-Shu,2014-Zhang-JFM,2015-Zhang-JCP}%
. Examples in this class are the discrete ordinate method \cite%
{2004-Li-JCP,2009-Li-JCP,2015-Li-PAS,2016-Shu-JCP}, the unified gas kinetic
scheme (UGKS) and the discrete UGKS \cite%
{2013-Shu-JCP,2014-Shu-JCP,2016-Shu-PRE,2016-Shu-JCP2,
2015-Guo-PRE,2016-Guo-PRE,2016-Xu-JCP,2017-Xu-JCP}, the regularized 13 (26)
moment approach \cite%
{2003-Struchtrup-POF,2004-Struchtrup-JFM,2007-PRL-Struchtrup,2017-Struchtrup-POF,2009-Gu-JFM}%
, the quadrature method of moments \cite%
{2008-Fox-JCP,2009-Fox-JCP,2012-Fox-JCP}, the lattice Boltzmann kinetic
method (LBKM) \cite{1992-BSV, 2001-Succi,2002-Succi,
1998-Wagner-PRL,1999-Wagner-PRE,2007-Yeomans-PRE,2016-Yeomans-PRL,2016-Wagner-PRE, 2005-Ansumali-PRL,2017-Ansumali-PRL,2017-Huang-JFM,2013-AG-EPL,2013-Shu-CF,2016-Shu-CMA,2014-AG-PRE,2015-AG-PRE,LiQing,2013-JSM}
or discrete Boltzmann method/model (DBM) \cite%
{Succi-DBM,2015-AG-SM,2016-AG-PRE,2016-AG-CF,2016-AG-CF2,2017-Lin-PRE,2017-Lin-SciRep,2018-Lin-CF}%
. In this paper, we focus only on the strategies for constructing
higher-order LBKM/DBM beyond NS hydrodynamics \cite%
{2007-Ansumali-PRL,2008-Ansumali-PRE,2017-Karlin-JFM,
2005-Succi-EPL,2005-Succi-POF,2006-Succi-EPL,
2015-Succi-PRE,2016-Succi-JSC,2016-Succi-PT,
2006-Shan-JFM,2011-Shan-PRE,2013-Zhang-JFM,
2005-Sofonea-PRE,2005-Sofonea-JCP,2009-Watari-PRE,
2014-JCP-Zhang,2006-Zhang-PRE,2008-Zhang-PRE}, that can be roughly
classified into the following five categories, i.e., the Hermite expansion
approach, the elaborate boundary condition way, the effective local
mean-free-path approach, the coupled particle-continuum scheme, and the
collisional lattice Boltzmann approach. The Hermite expansion approach,
presented by Shan \emph{et al.} \cite%
{2006-Shan-JFM,2011-Shan-PRE,2013-Zhang-JFM}, is a straightforward and
systematic framework for constructing higher-order LB approximations to the
Boltzmann-BGK equation beyond the NS level by using high-order Hermite
expansions with appropriate quadratures. In this way, the order of Hermite
expansion is responsible for obtaining correct kinetic moment relations.
Hence through incorporating higher-order terms in the Hermite expansions,
hydrodynamic models at various levels can be obtained at any order of
truncation of the Hermite polynomials. To capture the velocity-slip and
temperature-jump phenomena in the slip regime, an alternative way is to
design elaborate boundary conditions \cite%
{2005-Sofonea-PRE,2005-Sofonea-JCP,2009-Watari-PRE,2014-JCP-Zhang}, for
instance, the bounce back, specular reflection, diffuse-reflection, and
Maxwell-type boundary conditions, etc. In the third approach, Zhang \emph{et
al.} \cite{2006-Zhang-PRE,2008-Zhang-PRE} proposed a novel wall function to
modify the local mean-free-path and the relaxation time through which to
take into account the non-equilibrium characteristics in the Knudsen layer.
This simple treatment is effective for Knudsen numbers up to $1.58$. The
fourth approach \cite{2016-Succi-JSC,2016-Succi-PT} consists of two coupled
elements: the DSMC and LBKM which work not only for the weak non-equiulbrium
regions but also the strong non-equilibrium regions. The delicate
combination actually acts as an efficient multiscale strategy with respect
to the full DSMC. The last approach was presented by Green \emph{et al.}
\cite{2013-JSM}, the main difference between their method and the
conventional LBKM is the consideration of the detailed effects of
collisional interactions via the full collision operator of the Boltzmann
equation without any equilibrium based approximations. Such a treatment
makes the method particularly suitable for simulating highly non-equilibrium
flows with relative large Knudsen number, although it involves a greater
computational load due to the numerical solution of a system of coupled,
nonlinear ordinary differential equations when dealing with the five-fold
Boltzmann collision integral. Nevertheless, it should be noted that all the
above-mentioned attempts are suitable for isothermal or thermal case with
sufficiently small Mach number. Significant effort is still urgently
required to develop robust high-order LBKM/DBM for modeling highly
non-equilibrium flows with high Mach number and significant thermal effects.

To this end, we resort to DBM, which aims to probe the trans- and
supercritical fluid behaviors \cite{Succi-DBM} or to study simultaneously
the HNE and TNE behaviors, and has brought significant new physical insights
into the systems \cite%
{He-FOP,He-PRE,2015-AG-SM,2014-AG-PRE,2015-AG-PRE,2016-AG-PRE,2016-AG-CF,2016-AG-CF2,2017-Lin-PRE,2017-Lin-SciRep,2018-Lin-CF}.
Concretely, in this paper, we present a general framework for developing
trans-scale DBMs for high-speed compressible flows ranging from continuum to
transition regime through the construction of 
higher-order discrete equilibrium distribution function (DEDF) that
satisfies additional higher-order kinetic moments and the design of
higher-order isotropic discrete-velocity model (DVM) with smaller number of
discrete velocities; as a model example, we present a two-dimensional
compressible DBM with $26$ discrete velocities at the Burnett level;
determine the relations between macroscopic dissipations and non-equilibrium
measures defined through DBM, and provide a more general constitutive
relations for viscous stress and heat flux that can be used to improve
macroscopic modeling.

\section{Higher-order DBM and higher-order constitutive relations}

The key step of physical modeling of complex fluid system is the
coarse-graining process. The principle for such a simplification process is
that the physical quantities we choose to measure the system should stay
unchanged after simplification. On the whole, the discrete Boltzmann
trans-scale modeling procedure includes the following four steps, as shown
in Fig. 1: \newline
(I) Linearization of the collision term; \newline
(II) Discretization of the particle velocity space; \newline
(III) Determination of the required kinetic moments via Chapman-Enskog (CE)
analysis; \newline
(IV) Acquisition of the DEDF and DVM according to the required kinetic
moments.

\begin{figure}[tbp]
{\centerline{\epsfig{file=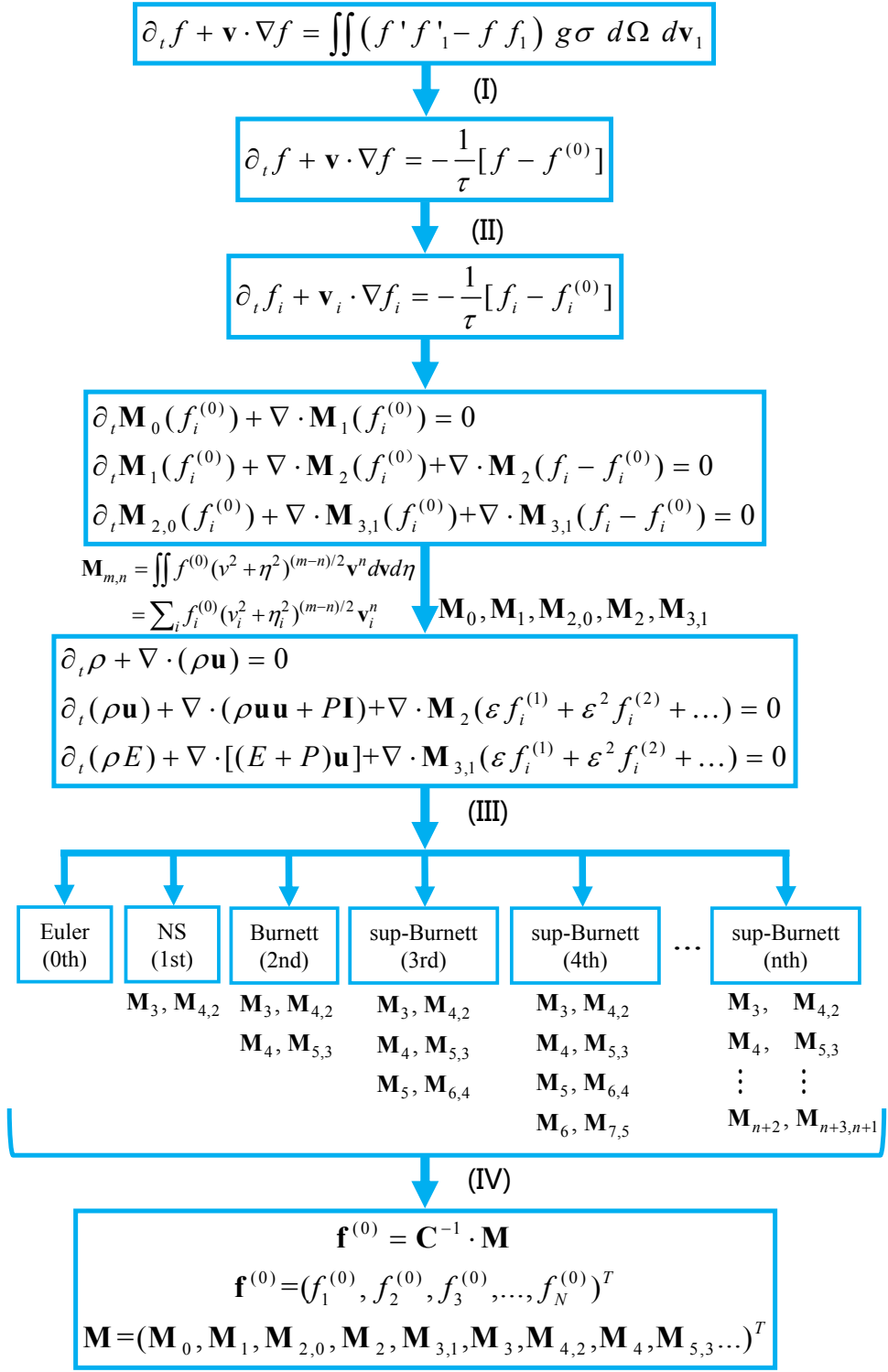,bbllx=2pt,bblly=3pt,bburx=351pt,bbury=541pt,
width=0.8\textwidth,clip=}}}
\caption{Flow chart for the discrete Boltzmann trans-scale modeling of
compressible flows.}
\end{figure}

Next, we explain what we really imply 
and what we conduct in each step. In step (I), it is well known that, the
original collision term of the Boltzmann equation $J(f,f^*)$ is too complex to
be solved directly, where $f$ and $f^*$ are distribution functions before and after collisions, respectively.
The simplest way to linearize it, is to introduce a
local equilibrium distribution function $f^{(0)}$ and write the collision
term into the BGK-like form \cite{BGK} $J=-\frac{1}{ \tau}[{f}-{f}^{(0)}]$,
where ${{f}^{(0)}}={{\frac{{\rho }}{2\pi RT}}}{{\ \left( \frac{{1}}{2\pi nRT}%
\right) }^{1/2}}\exp \left[ -\frac{{{\left( \mathbf{v}-\mathbf{u}\right) }%
^{2}}}{2RT}-\frac{{{\eta }^{2}}}{2nRT}\right] $ is the Maxwellian
distribution function with $\rho$, $\mathbf{v}$, $\mathbf{u}$, $T$ are the
local density, particle velocity, flow velocity and temperature,
respectively. $R$ is the gas constant, $\eta$ is a free parameter introduced
to describe the $n$ extra degrees of freedom corresponding to molecular
rotation and/or vibration. Owing to its simplicity, the BGK approximation is
the most extensively used. Besides this, other models including the
ellipsoidal statistical BGK model \cite{ES}, Shakhov model \cite{Shakhov},
Rykov model \cite{Rykov}, and Liu model \cite{Liu1990}, etc., have also been
used to simplify the full collision operator of the Boltzmann equation and
to tune the Prandtl number.

To perform simulation, we have to write the BGK-like Boltzmann equation in a
discrete form, i.e., the discrete Boltzmann equation
\begin{equation}
\partial _{t}f_{i}+\mathbf{v}_{i}\cdot \bm{\nabla}f_{i}=-\frac{1}{\tau }%
[f_{i}-f_{i}^{(0)}],  \label{DB}
\end{equation}%
which results in the second step. 
The discretization of six-dimensional phase-space, i.e.
position-and-velocity space, is however prohibitively expensive from the
computational standpoint. To find an effective way to discretize the
particle velocity space, we go back to consider what we really need and at
which level the hydrodynamic equations are recovered from the discrete
Boltzmann equation. In fact, in the DB modeling, we do not expect to
describe the system by using specific values of the discrete distribution
function $f_{i}$, but rather the kinetic moments of $f_{i}$. The
CE analysis informs us that the calculations of all the kinetic
moments of $f_{i}$ can finally resort to those of the DEDF $f_{i}^{(0)}$.
Therefore, we should 
ensure that these kinetic moments of $f^{(0)}$, originally in integral form,
can be calculated in summation form during the modeling process.

To determine which level the hydrodynamic equations are recovered and which
kinetic moments of $f_{i}^{(0)}$ are needed, one can derive the hydrodynamic
equations from the discrete Boltzmann equation via CE
multiscale expansion. Essentially, the derivation of hydrodynamic equations
from discrete Boltzmann equation is sufficient but not necessary. Compared
with the traditional modeling schemes based on continuum assumption, DBM is
a kind of different scheme to model the non-equilibrium flows. DBM modeling
and simulation do not need the hydrodynamic equations; one needs only to
determine the required kinetic moments via CE procedure, which is one of the
prominent advantages of DBM and the key point of the manuscript. Then we
show the derivation from discrete Boltzmann equation to hydrodynamic
equations, which serves the purpose of showing why such a modeling scheme
does work.

It is found that, when $f_{i}^{(0)}$ satisfies the following five kinetic
moments,
\begin{equation}
\mathbf{M}_{0}=\sum\nolimits_{i}f_{i}^{(0)}=\rho \text{,}  \label{M0}
\end{equation}%
\begin{equation}
\mathbf{M}_{1}=\sum\nolimits_{i}f_{i}^{(0)}\mathbf{v}_{i}=\rho \mathbf{u}%
\text{,}  \label{M1}
\end{equation}%
\begin{equation}
\mathbf{M}_{2,0}=\sum\nolimits_{i}\frac{1}{2}f_{i}^{(0)}(v_{i}^{2}+\eta
_{i}^{2})=\frac{1}{2}\rho \lbrack (n+2)RT+u^{2}]\text{,}  \label{M20}
\end{equation}%
\begin{equation}
\mathbf{M}_{2}=\sum\nolimits_{i}f_{i}^{(0)}\mathbf{v}_{i}\mathbf{v}_{i}=\rho
(RT\mathbf{I}+\mathbf{uu})\text{,}  \label{M2}
\end{equation}%
\begin{equation}
\mathbf{M}_{3,1}=\sum\nolimits_{i}\frac{1}{2}f_{i}^{(0)}(v_{i}^{2}+\eta
_{i}^{2})\mathbf{v}_{i}=\frac{1}{2}\rho \mathbf{u}[(n+4)RT+u^{2}]\text{,}
\label{M31}
\end{equation}%
taking moments of Eq. (\ref{DB}) with the collision invariant vector $1$, $%
\mathbf{v}_{i}$, $\frac{1}{2}(\mathbf{v}_{i}^{2}+\eta _{i}^{2})$, gives rise
to the following generalized set of thermohydrodynamic equations
\begin{equation}
\partial _{t}\rho +\bm{\nabla}\cdot (\rho \mathbf{u})=0\text{,}  \label{1}
\end{equation}%
\begin{equation}
\partial _{t}(\rho \mathbf{u})+\bm{\nabla}\cdot (\rho \mathbf{uu}+P\mathbf{I}%
+\bm{\Delta}_{2}^{\ast })=0\text{,}  \label{2}
\end{equation}%
\begin{equation}
\partial _{t}(\rho E)+\bm{\nabla}\cdot \lbrack (E+P)\mathbf{u}+\bm{\Delta}%
_{2}^{\ast }\cdot \mathbf{u}+\bm{\Delta}_{3,1}^{\ast }]=0\text{,}  \label{3}
\end{equation}%
where $P=\rho RT$ is the local hydrostatic pressure and $E=c_{v}T+u^{2}/2$
the total energy with $c_{v}=(n+2)R/2$ the specific heat at constant volume.
Here \textquotedblleft satisfaction" means the moments calculated from the
summation of ${f}_{i}^{(0)}$ should be the same as those from the
integration of ${f}^{(0)}$
\begin{equation}
\sum\nolimits_{i}{f_{i}^{(0)}\mathbf{\Psi }(}\mathbf{v}_{i},\eta _{i})={%
\mathbf{M}}_{m,n}={\iint {{{f}^{(0)}}}\mathbf{\Psi }}(\mathbf{v,}\eta {)}d%
\mathbf{v}d\eta \text{,}
\end{equation}%
where ${\mathbf{\Psi }}(\mathbf{v}_{i},\eta _{i})=[1,\mathbf{v}_{i},\frac{1}{%
2}(v_{i}^{2}+\eta _{i}^{2}),\mathbf{v}_{i}\mathbf{v}_{i},\frac{1}{2}%
(v_{i}^{2}+\eta _{i}^{2})\mathbf{v}_{i}]^{T}$. Note that Eqs. (\ref{2})-(\ref%
{3}) are unclosed. To close these equations at various levels, we should
deduce the explicit expressions for the TNE measures $\bm{\Delta }_{2}^{\ast
}$ and $\bm{\Delta}_{3,1}^{\ast }$. Physically, these two measures reflect
molecular individualism on top of organized collective motion, which are
conventionally labeled as \emph{non-organised moment fluxes} (NOMF)
\begin{equation}
\bm{\Delta}_{2}^{\ast }=\mathbf{M}_{2}^{\ast
}(f_{i}-f_{i}^{(0)})=\sum\nolimits_{i}(f_{i}-f_{i}^{(0)})\mathbf{v}%
_{i}^{\ast }\mathbf{v}_{i}^{\ast },
\end{equation}%
and \emph{non-organised energy fluxes} (NOEF),
\begin{equation}
\bm{\Delta }_{3,1}^{\ast }=\mathbf{M}_{3,1}^{\ast
}(f_{i}-f_{i}^{(0)})=\sum\nolimits_{i}(f_{i}-f_{i}^{(0)})\frac{v_{i}^{\ast
2}+\eta _{i}^{2}}{2}\mathbf{v}_{i}^{\ast },
\end{equation}%
respectively. $\mathbf{M}_{2}^{\ast }$ and $\mathbf{M}_{3,1}^{\ast }$ are
kinetic central moments. Compared with NS and Burnett equations, $\bm{\Delta}%
_{2}^{\ast }$ ($\bm{\Delta }_{3,1}^{\ast }$) corresponds to the full viscous
stress tensor $\bm{\sigma}$ (heat flux $\mathbf{j}_{q}$). Therefore, the
relation between TNE measure and macroscopic dissipation is clarified. Of
course, the derivations of $\bm{\Delta}_{2}^{\ast }$ and $\bm{\Delta }%
_{3,1}^{\ast }$ will induce additional requirements on moments of $%
f_{i}^{(0)}$.

Step III demonstrates that to recover hydrodynamic equations at different
levels, $f_{i}^{(0)}$ should satisfy different additional kinetic moments.
The requirements on kinetic moments of $f_{i}^{(0)}$ can be determined as
follows. To perform the CE expansion on both sides of Eq. ( \ref%
{DB}), we first introduce expansions
\begin{equation}
f_{i}=f_{i}^{(0)}+\epsilon f_{i}^{(1)}+\epsilon ^{2}f_{i}^{(2)}+\cdots \text{%
,}  \label{CE1}
\end{equation}
\begin{equation}
\partial _{t}=\epsilon \partial _{t_{1}}+\epsilon ^{2}\partial
_{t_{2}}+\cdots \text{,}  \label{CE2}
\end{equation}%
\begin{equation}
\bm{\nabla}=\epsilon \bm{\nabla}_{1}\text{,}  \label{CE3}
\end{equation}%
where $\epsilon ^{j}f_{i}^{(j)}$ is the $j$-th order departure from $%
f_{i}^{(0)}$ in Knudsen number $\epsilon $, and $\epsilon ^{j}\partial
_{t_{j}}$ is $j$-th order term in $\epsilon $.  Substituting Eqs. (\ref{CE1}%
)-(\ref{CE3}) into Eq. (\ref{DB}) and equating terms that have the same
orders in $\epsilon $ gives the following formulations for $f_{i}^{(1)}$ and
$f_{i}^{(2)}$,
\begin{equation}
\epsilon f_{i}^{{(1)}}=-\tau \lbrack \epsilon {\partial }%
_{t_{1}}f_{i}^{(0)}+\epsilon \bm{\nabla}_{1}\cdot (f_{i}^{(0)}\mathbf{v}%
_{i})],
\end{equation}%
and
\begin{eqnarray}
\epsilon ^{2}f_{i}^{{(2)}} &=&-\tau \lbrack \epsilon ^{2}{\partial }%
_{t_{2}}f_{i}^{(0)}{+\epsilon \partial }_{t_{1}}(\epsilon
f_{i}^{(1)})+\epsilon \bm{\nabla}_{1}\cdot (\epsilon f_{i}^{(1)}\mathbf{v}%
_{i})]  \notag \\
&=&-\tau \epsilon ^{2}{\partial }_{t_{2}}f_{i}^{(0)}{+\tau }^{2}\epsilon ^{2}%
{\partial }_{t_{1}}^{2}f_{i}^{(0)}{+{\tau }^{2}\epsilon ^{2}{\partial }%
_{t_{1}}[\bm{\nabla}_{1}\cdot (f_{i}^{(0)}\mathbf{v}}_{i}{)]}  \notag \\
&&+\tau ^{2}\epsilon ^{2}\bm{\nabla}_{1}\cdot \lbrack {\partial }_{t_{1}}{%
f_{i}^{(0)}}\mathbf{v}_{i}+\bm{\nabla}_{1}\cdot (f_{i}^{(0)}\mathbf{v}_{i}%
\mathbf{v}_{i})]\text{.}
\end{eqnarray} 
It is clear that (i) $\epsilon f_{i}^{(1)}$ and $\epsilon ^{2}f_{i}^{(2)}$ 
can be expressed as formulations of $f_{i}^{(0)}$; (ii) $\epsilon f_{i}^{(1)}
$  includes a polynomial of $\mathbf{v}_{i}$ of one order higher than that in
$f_{i}^{(0)}$; (iii)  $\epsilon ^{2}f_{i}^{(2)}$  includes a polynomial of $%
\mathbf{v}_{i}$ of one order higher than that in  $\epsilon f_{i}^{(1)}$  but
two orders higher than that in $f_{i}^{(0)}$. Obviously, to achieve explicit
expressions for the first-order constitutive relations, $\bm{\Delta}%
_{2}^{(1)\ast }=\sum\nolimits_{i}\epsilon f_{i}^{(1)}\mathbf{v}_{i}^{\ast }%
\mathbf{v}_{i}^{\ast }$  and  $\bm{\Delta }_{3,1}^{(1)\ast
}=\sum\nolimits_{i}\epsilon f_{i}^{(1)}\frac{v_{i}^{\ast 2}+\eta _{i}^{2}}{2}%
\mathbf{v}_{i}^{\ast }$ , the highest order non-equilibrium kinetic moments
that $f_{i}^{(0)}$ should further satisfy are
\begin{equation}
{\mathbf{M}}_{3}=\sum\nolimits_{i}{f_{i}^{(0)}}\mathbf{v}_{i}\mathbf{v}_{i}%
\mathbf{v}_{i}=\rho (RT\bm{\Theta}+\mathbf{uuu})\text{,}  \label{M3}
\end{equation}%
\begin{equation}
{\mathbf{M}}_{4,2}=\sum\nolimits_{i}{f_{i}^{(0)}}\frac{v_{i}^{2}+\eta
_{i}^{2}}{2}\mathbf{v}_{i}\mathbf{v}_{i}=\rho \lbrack (\frac{n+4}{2}RT+\frac{%
u^{2}}{2})RT\mathbf{I}+(\frac{n+6}{2}RT+\frac{u^{2}}{2})\mathbf{uu}]\text{,}
\label{M42}
\end{equation}%
respectively. Similarly, to achieve explicit expressions for the
second-order constitutive relations, $f_{i}^{(0)}$ should satisfy $\mathbf{M}%
_{4}$ and $\mathbf{M}_{5,3}$,
\begin{equation}
\mathbf{M}_{4}=\sum\nolimits_{i}f_{i}^{(0)}\mathbf{v}_{i}\mathbf{v}_{i}%
\mathbf{v}_{i}\mathbf{v}_{i}=\rho (R^{2}T^{2}\bm{\Pi }+RT\bm{\Xi}+\mathbf{%
uuuu})\text{,}  \label{M4}
\end{equation}%
\begin{equation}
\mathbf{M}_{5,3}=\sum\nolimits_{i}\frac{1}{2}f_{i}^{(0)}(v_{i}^{2}+\eta
_{i}^{2})\mathbf{v}_{i}\mathbf{v}_{i}\mathbf{v}_{i}=\rho \lbrack (\frac{n+8}{%
2}RT+\frac{u^{2}}{2})\mathbf{uuu}+(\frac{n+6}{2}RT+\frac{u^{2}}{2})RT%
 \bm{\Theta} ]\text{,}  \label{M53}
\end{equation}%
with $ \bm{\Theta}=(u_{\alpha }\delta _{\beta \gamma }+u_{\beta
}\delta _{\alpha \gamma }+u_{\gamma }\delta _{\alpha \beta })\widehat{%
\mathbf{e}}_{\alpha }\widehat{\mathbf{e}}_{\beta }\widehat{\mathbf{e}}%
_{\gamma }$, $\bm{\Pi }=(\delta _{\alpha \beta }\delta _{\gamma \lambda
}+\delta _{\alpha \gamma }\delta _{\beta \lambda }+\delta _{\alpha \lambda
}\delta _{\beta \gamma })\widehat{\mathbf{e}}_{\alpha }\widehat{\mathbf{e}}%
_{\beta }\widehat{\mathbf{e}}_{\gamma }\widehat{\mathbf{e}}_{\lambda }$, $%
\bm{\Xi }=(u_{\alpha }u_{\beta }\delta _{\gamma \lambda }+u_{\alpha
}u_{\gamma }\delta _{\beta \lambda }+u_{\alpha }u_{\lambda }\delta _{\beta
\gamma }+u_{\beta }u_{\gamma }\delta _{\alpha \lambda }+u_{\beta }u_{\lambda
}\delta _{\alpha \gamma }+u_{\gamma }u_{\lambda }\delta _{\alpha \beta })%
\widehat{\mathbf{e}}_{\alpha }\widehat{\mathbf{e}}_{\beta }\widehat{\mathbf{e%
}}_{\gamma }\widehat{\mathbf{e}}_{\lambda }$, $(\widehat{\mathbf{e}}_{\alpha
},\widehat{\mathbf{e}}_{\beta },\widehat{\mathbf{e}}_{\gamma },\widehat{%
\mathbf{e}}_{\lambda })$ denote unit vectors along the $\alpha $, $\beta $, $%
\gamma $ and $\lambda $ axes of a fixed coordinate system.

By using the above needed kinetic moments and after some tedious but
straightforward algebraic manipulation, we acquire relations between
thermodynamic forces and fluxes,
\begin{equation}
\bm{\Delta }_{2}^{\ast (1)}=-\mu \lbrack \bm{\nabla}\mathbf{u}+(\bm{\nabla}%
\mathbf{u})^{T}-\frac{2}{n+2}\mathbf{I}\bm{\nabla}\cdot \mathbf{u}]=-%
\bm{\sigma }_{\text{NS}},  \label{vis1}
\end{equation}%
\begin{equation}
\bm{\Delta }_{3,1}^{\ast (1)}=-\kappa \bm{\nabla}T=-\mathbf{j}_{q,\text{NS}},
\label{heat1}
\end{equation}%
where the first-order NOMF and NOEF are just the negative viscous stress
tensor and heat flux at the NS level, with $\mu =P\tau $, $\kappa =$ $%
c_{p}P\tau $ are viscosity coefficient and heat conductivity, respectively.
Here $c_{p}=(n+4)R/2$ is the specific heat at constant pressure. Expressions
for the second-order constitutive relations,  $\bm{\Delta }_{2}^{\ast
(2)}=\sum\nolimits_{i}\epsilon ^{2}f_{i}^{(2)}\mathbf{v}_{i}^{\ast }\mathbf{v%
}_{i}^{\ast } =-(\bm{\sigma }_{\text{Burnett}}-\bm{\sigma}_{\text{NS}})$, $%
\bm{\Delta }_{3,1}^{\ast (2)}=\sum\nolimits_{i}\epsilon ^{2}f_{i}^{(2)}\frac{%
v_{i}^{\ast 2}+\eta _{i}^{2}}{2}\mathbf{v}_{i}^{\ast }=-(\mathbf{j}_{q,\text{%
Burmett}}-\mathbf{j}_{q,\text{NS}})$  are displayed in the Appendix. So far,
the higher-order constitutive relations for viscous stress and heat transfer
at the Burnett level have been given by $\bm{\Delta }_{2}^{\ast }=%
\bm{\Delta}_{2}^{\ast (1)}+\bm{\Delta }_{2}^{\ast (2)}$ and $\bm{\Delta }_{3,1}^{\ast
}=\bm{\Delta
}_{3,1}^{\ast (1)}+\bm{\Delta }_{3,1}^{\ast (2)}$, which are expected to
noticeably improve the macroscopic modeling. Counterparts at super-Burnett
levels can be deduced in a similar way.

\begin{figure}[tbp]
{%
\centerline{\epsfig{file=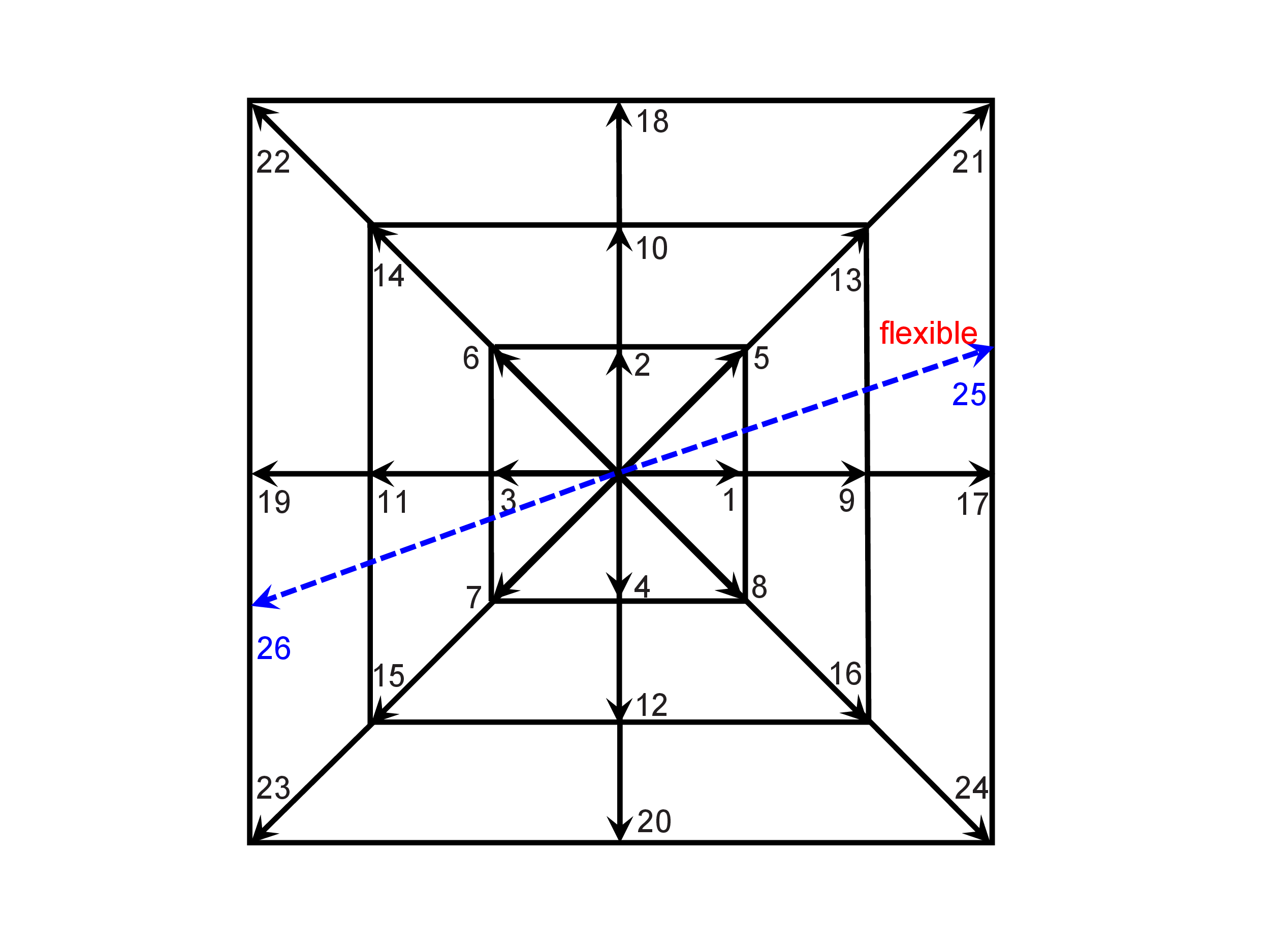,bbllx=135pt,bblly=56pt,bburx=570pt,bbury=491pt,
width=0.45\textwidth,clip=}}}
\caption{Schematic of the D2V26 discrete-velocity model, where $\mathbf{v }%
_{25}$(=$-\mathbf{v}_{26}$) is a flexible vector, adjusted to guarantee the
existence of $\mathbf{C}^{-1}$.}
\end{figure}

Finally, in step IV, we obtain the analytical expression for DEDF via
inversely solving the required kinetic moments. Details are as follows. In
the two-dimensional case, the above moments $\mathbf{M}_{0}$, $\mathbf{M}%
_{1} $, $\mathbf{M}_{2,0}$, $\mathbf{M}_{2}$, $\mathbf{M}_{3,1}$, $\mathbf{M}%
_{3}$, $\mathbf{M}_{4,2}$, $\mathbf{M}_{4}$ and $\mathbf{M}_{5,3}$ have $25$
components. For physical symmetry and numerical stability, we add the
following one
\begin{equation}
\mathbf{M}_{4,0}=\sum\nolimits_{i}\frac{1}{2}f_{i}^{(0)}(v_{i}^{2}+\eta
_{i}^{2})^{2}=\rho \lbrack \frac{3n^{2}+4n+8}{2}R^{2}T^{2}+(n+4)RTu^{2}+%
\frac{u^{4}}{2}]\text{.}  \label{M40}
\end{equation}%
These moments can be rewritten in a matrix form
\begin{equation}
\mathbf{C} \cdot \mathbf{f}^{(0)}=\mathbf{M}\text{,}  \label{Feq_1}
\end{equation}%
where $\mathbf{f}^{(0)}=(f_{1}^{(0)},f_{2}^{(0)},\cdots ,f_{26}^{(0)})^{T}$,
$\mathbf{M}=(M_{1},M_{2},\cdots
,M_{26})^{T}=(M_{0},M_{1x},M_{1y},...,M_{4,0})^{T}$ is the set of moments of
$f_{i}^{(0)}$. $\mathbf{C}=(\mathbf{c}_{1},\mathbf{c}_{2},...,\mathbf{c}%
_{26})$ is a $26\times 26$ matrix bridging the DEDF and the kinetic moments
with $\mathbf{c}_{i}=(1,v_{ix},v_{iy},...,\frac{1}{2}(\bm{v}_{i}^{2}+\eta
_{i}^{2})^{2})^{T}$. As a result, $\mathbf{f}^{(0)}$ can be calculated in
the following way \cite{2013-AG-EPL},
\begin{equation}
\mathbf{f}^{(0)}=\mathbf{C}{^{-1}} \cdot \mathbf{M}\text{,}
\end{equation}%
where $\mathbf{C}{^{-1}}$ is the inverse of matrix $\mathbf{C}$. A
two-dimensional DVM with 26 discrete velocities, schematically drawn in
Fig.2, is appropriately designed to discretize the velocity space and to
ensure the existence of $\mathbf{C}^{-1}$
\begin{equation}
(v_{ix},v_{iy})=\left\{
\begin{array}{cc}
\text{cyc}:c(\pm 1,0) & \text{for}\quad 1\leq i\leq 4 \\
c(\pm 1,\pm 1) & \text{for}\quad 5\leq i\leq 8 \\
\text{cyc}:2c(\pm 1,0) & \ \text{for}\quad 9\leq i\leq 12 \\
2c(\pm 1,\pm 1) & \ \quad \text{for}\quad 13\leq i\leq 16 \\
\text{cyc}:3c(\pm 1,0) & \quad \ \text{for}\quad 17\leq i\leq 20 \\
3c(\pm 1,\pm 1) & \quad \ \text{for}\quad 21\leq i\leq 24 \\
c(3,1),-c(3,1) & \quad \ \text{for}\quad 25\leq i\leq 26%
\end{array}%
\right. ,
\end{equation}
where ``cyc" indicates the cyclic permutation. For $1\leq
i\leq 4$, $\eta _{i}=i\eta _{0}$; \bigskip for $5\leq i\leq 8$, $\eta
_{i}=(i-4)\eta _{0}$; otherwise $\eta _{i}=0$. The choosing of $\mathbf{v}%
_{25}$ is flexible as long as $\mathbf{v}_{25}=-\mathbf{v}_{26}$, where $c$
and $\eta _{0}$ are two free parameters, adjusted to guarantee the existence
of $\mathbf{C}^{-1}$ and to optimize the properties of the model. The
specific-heat ratio can be defined as $\gamma =c_{p}/c_{v}=(n+4)/(n+2)$.

After the accomplishment of physical modeling, we solve Eq. (\ref{DB}) to
update $f_{i}$ via finite difference schemes. Physical quantities, such as
density, velocity, temperature, pressure, viscous stress and heat flux are
calculated from kinetic moments of $f_{i}$ and equation of state: $\rho
=\sum\nolimits_{i}f_{i}$, $\mathbf{u}=\sum\nolimits_{i}f_{i}\mathbf{v}%
_{i}/\rho $, $T=\frac{1}{(n+2)R}[\sum\nolimits_{i}f_{i}(v_{i}^{2}+\eta
_{i}^{2})/\rho -u^{2}]$, $P=\rho RT$, $\bm{\Delta}_{2}^{\ast
}=\sum\nolimits_{i}(f_{i}-f_{i}^{(0)})\mathbf{v}_{i}^{\ast }\mathbf{v}%
_{i}^{\ast }$ and $\bm{\Delta}_{3,1}^{\ast
}=\sum\nolimits_{i}(f_{i}-f_{i}^{(0)})\frac{v_{i}^{\ast 2}+\eta _{i}^{2}}{2}%
\mathbf{v}_{i}^{\ast }$.

It is noteworthy that (a) the approach for calculating DEDF is general,
straightforward, and independent of the Gaussian quadrature formula; (b) the
number of discrete velocities used here can be as small as that of the
independent kinetic moment relations. Compared with other kinetic methods,
DBM adapts the minimal set of discrete velocities and consequently it enjoys
a high computational efficiency; (c) the model casts off the standard
``propagation + collision" mode and frees from the combination of spatial
and temporal discretizations. The sets of particle velocities are highly
flexible in magnitude and number,
which substantially improves the numerical stability, and consequently, is
much more convenient to meet the requirements for simulating compressible
flows; (d) to access the behavior of the system farther away from
equilibrium, one needs 
to add more kinetic moment relations into ${\mathbf{\Psi }}(\mathbf{v}%
_{i},\eta _{i})$. Then ${\mathbf{\Psi }}(\mathbf{v}_{i},\eta _{i})$ owns
more elements and $f_{i}^{(0)}$ becomes more complicated, 
and more discrete velocities are needed. Compared with the corresponding
hydrodynamic equations whose complexity will sharply increases with
increasing the degree of TNE effects, the modeling process of DBM
is only mildly affected. This is a major benefit of the discrete velocity
representation versus the hierarchical Hermite expansion, which generates
highly non-linear tensors at each increasing order.
(e) being able to recover the NS (Burnett) model is only one of the
functions of the DBM. The DBM presents a kind of new approach and a set of
convenient and efficient tools to describe, measure and analyze the
non-equilibrium behaviors, by calculating the difference between kinetic
moments of discrete distribution functions and DEDF, $\bm{\Delta}_{m}=%
\mathbf{M}_{m}(f-f^{(0)})$ and $\bm{\Delta }_{m}^{\ast }=\mathbf{M}%
_{m}^{\ast }(f-f^{(0)})$. From this point of view, a DBM is roughly
equivalent to a hydrodynamic model supplemented by a coarse grained model of
the TNE effects. (f) at last, we stress that, via the DBM, it is
straightforward to perform multi-scale simulations over a wide range of
Knudsen number by switching the effective parameter controlling the TNE
extent. This is because the multiscale modeling of DBM is under the same
framework without message passing between models at different scales. These
outstanding advantages make DBM a particularly appealing methodology for
investigating non-equilibrium flows.

Meanwhile, we point out that, owing to the utilization of a single
relaxation time in the collision term, the Prandtl number becomes fixed at a
specific value $\Pr =1$. To overcome this limitation, one convenient way is
to add an external forcing term $I_i$ into the right-hand-side of Eq. (\ref%
{DB}) to modify the BGK collision operator \cite{2012-Gan-CTP}, $%
I_{i}=[ART+B(\mathbf{v}_{i}-\mathbf{u})^{2}]f_{i}^{(0)}$ with $A=-2B$ and $B=%
\frac{1}{2\rho T^{2}}\bm{\nabla}\cdot \lbrack \frac{4+n}{2}\rho Tq\bm{\nabla}%
T]$. As a result, the heat conductivity has been changed to be $\kappa
=c_{p}P(\tau +q)$, and the Prandtl number $\Pr =\frac{\tau }{\tau +q}$.
Besides its conciseness, more importantly, this approach does not give rise
to additional kinetic moments requirement.

\section{Numerical Simulations and Analysis}

In this section, several typical benchmarks, ranging from subsonic to
hypersonic, are conducted to validate the model. Afterwards, we investigate
carefully the performances of the new model for describing compressible
flows over a wide range of Knudsen numbers. To improve the numerical
stability, efficiency, and to accurately capture the complex characteristic
structures, the third-order implicit-explicit Runge-Kutta finite difference
scheme \cite{IMEX} is adopted to discretize the temporal derivative; the
second-order non-oscillatory non-free-parameter and dissipative finite
difference (FD) scheme is used to discretize the spatial derivatives for the
second and third Riemann problems; for other problems considered, the
fifth-order weighted essentially nonoscillatory FD scheme is applied to
calculate the spatial derivatives. The adoption of the FD scheme makes the
boundary condition (BC) easily incorporated into the model, which is exactly
the same as that implemented in the conventional computational fluid
dynamics (CFD) methods. The discrete Boltzmann equation, particle velocity,
and hydrodynamic quantities have been nondimensionalized by suitable
reference variables \cite{2011-Gan-PRE}. Among which, three independent ones
are the characteristic flow length scale $L_{0}$, the reference density ${{%
\rho }_{0}}$ and the reference temperature $T_{0}$. The other characteristic
variables are defined as $u_{0}=\sqrt{RT_{0}}$, $t_{0}=L_{0}/u_{0}$, and $%
P_{0}={{\rho }_{0}RT}_{0}$. In our simulations, we assume that the fluid is
air under normal conditions, then the scales used to specify the magnitudes
of the density, temperature, fluid velocity components\textit{\ }are ${{\rho
}_{0}=1.165}${kg/m}$^{3}$, $T_{0}=303$K, and $u_{0}=\sqrt{RT_{0}}\approx
294.892$m/s with $R=287$J/(kg$\cdot$ K), respectively.

\subsection{Riemann Problems}

\subsubsection{Sod shock tube}

\begin{figure}[tbp]
{%
\centerline{\epsfig{file=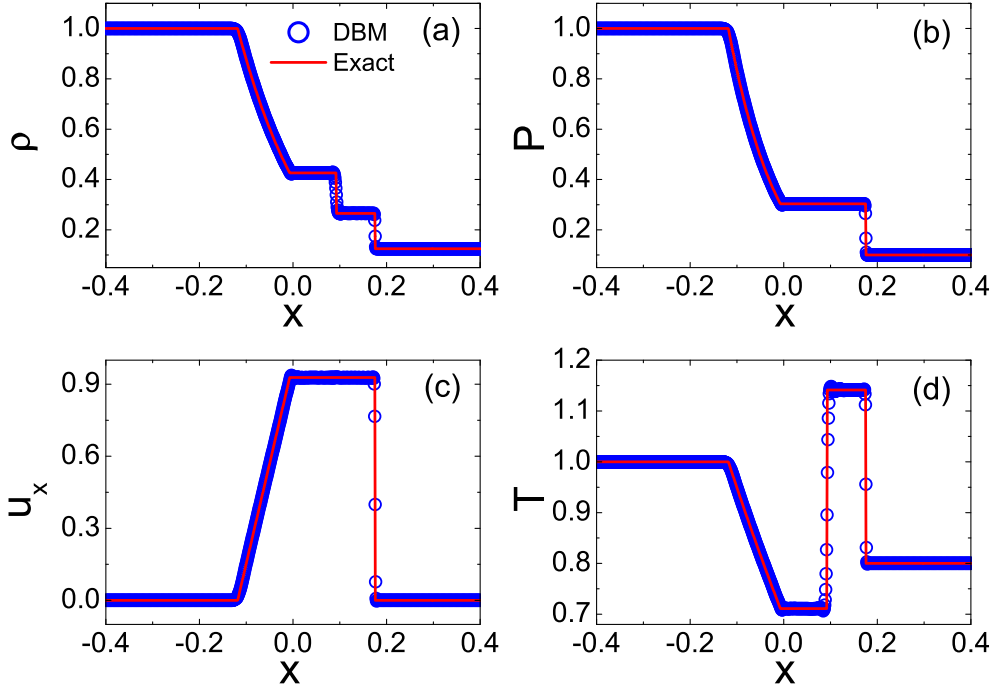,bbllx=7pt,bblly=3pt,bburx=576pt,bbury=434pt,
width=0.8\textwidth,clip=}}}
\caption{Comparisons between DBM simulations and the exact solutions for the
Sod shock tube, where $t=0.1$ and $\protect\gamma=1.4$. (a) Density, (b)
pressure, (c) velocity, and (d) temperature.}
\end{figure}

The first test case is the standard Sod shock problem with the following
initial conditions
\begin{equation}
\left\{
\begin{array}{c}
(\rho ,T,u_{x},u_{y})|_{L}=(1.0,1.0,0.0,0.0), \\
(\rho ,T,u_{x},u_{y})|_{R}=(0.125,0.8,0.0,0.0),%
\end{array}%
\right.
\end{equation}%
where subscripts \textquotedblleft L" and \textquotedblleft R" stand for
macroscopic variables at the left and right sides of the discontinuity. It
is a classical test in the study of compressible flows which consists of (i)
a shock wave propagating into the low pressure region, (ii) a rarefaction
wave expanding into the high pressure part, and (iii) a contact
discontinuity moving rightward. The periodic BC is imposed in the $y$
direction. In the $x$ direction, we apply the supersonic inflow BC \cite%
{QuKun,HYL}, i.e., $f_{i,-2,t}=f_{i,-1,t}=f_{i,0,t}=f_{i,1,t=0}^{(0)}$,
where $-2$, $-1$, and $0$ are indexes of three ghost nodes out of the left
boundary. Such a BC means that the system at the boundary stays as their
corresponding equilibrium state, or in other words, the macroscopic
quantities on the boundary nodes keep at their initial values $(\rho ,%
\mathbf{u},T)_{-2,t}=(\rho ,\mathbf{u},T)_{-1,t}=(\rho ,\mathbf{u}%
,T)_{0,t}=(\rho ,\mathbf{u},T)_{1,t=0}$. On the right side, we can operate
in a similar way. BC implemented on the distribution function and
macroscopic quantities may be referred to as the mesoscopic BC and the
macroscopic BC, respectively, which are consistent with each other.
Moreover, when the external environment is out-of-equilibrium, the
non-equilibrium part $f_{i,I}^{(\text{neq})}$ can be obtained from the inner
lattice nodes via the extrapolation method, which is a merit of DBM over the
traditional CFD. BCs for the following test cases are consistent with what
we imposed above. Parameters are set to be $\Delta x=\Delta y=10^{-3}$ , $%
\Delta t=10^{-4}$, $\tau =10^{-5}$, $c=1$, $\eta _{0}=1.5$, and $\gamma =1.4$%
. The lattice points are $2000\times 4$. Figure 3 exhibits the computed
density, pressure, velocity, and temperature profiles at $t=0.1$, where
circles indicate results from DBM simulations and solid lines from Riemann
solutions. The two sets of results coincide with each other. Moreover, the
shock wave, expanding wave and contact discontinuity are well captured with
severely curtailed numerical dissipation and effectively refrained
unphysical oscillations.

\subsubsection{Modified Lax shock tube}

\begin{figure}[tbp]
{%
\centerline{\epsfig{file=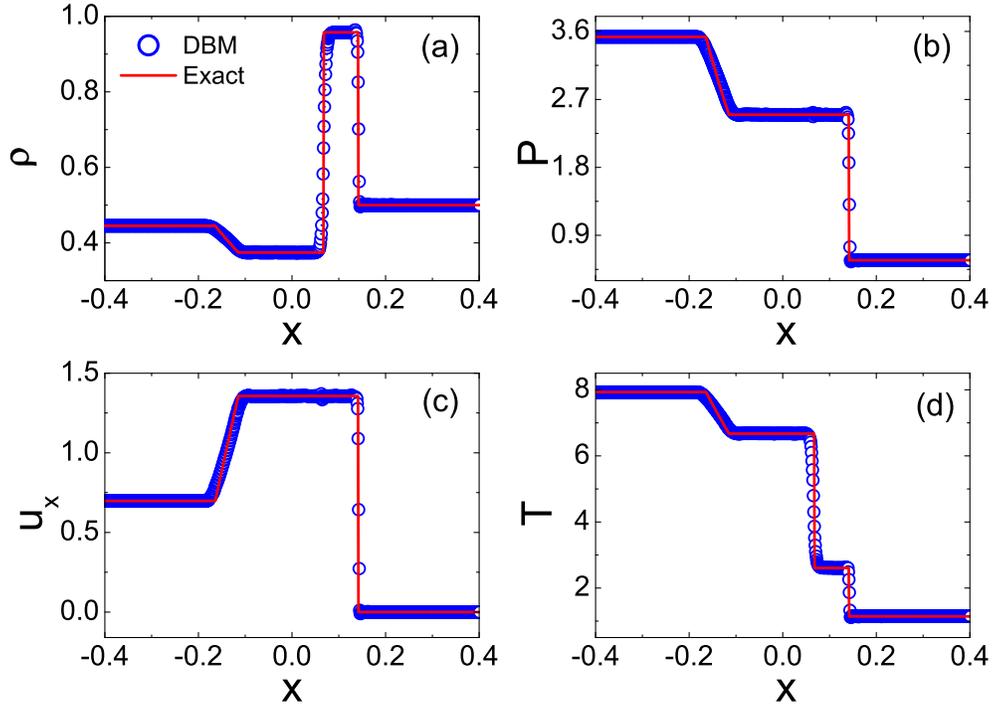,bbllx=7pt,bblly=3pt,bburx=576pt,bbury=434pt,
width=0.8\textwidth,clip=}}}
\caption{Comparisons between DBM simulations and the exact solutions for the
Lax shock tube, where $t=0.07$ and $\protect\gamma=2$. (a) Density, (b)
pressure, (c) velocity, and (d) temperature.}
\end{figure}

To further highlight robustness of the model, we construct a modified Lax
shock tube with larger velocity difference
\begin{equation}
\left\{
\begin{array}{c}
(\rho ,T,u_{x},u_{y})|_{L}=(0.445,7.928,0.698,0.0), \\
(\rho ,T,u_{x},u_{y})|_{R}=(0.50,1.142,-0.698,0.0).%
\end{array}%
\right.
\end{equation}
Comparisons between simulation results and the exact solutions at $t=0.07$
are plotted in Fig. 4, where $c=1.7$, $\eta_{0}=6.0$, and $\gamma=2$, other
parameters are unchanged. The two sets of results agree excellently with
each other. Additionally, the shock wave and contact discontinuity are
captured stably and no overshoots nor spurious oscillations appear.
Enlargement of the shock wave parts shows that it spreads over three to four
grid cells, demonstrating that the present model has a high resolving power
in capturing such complex structure.

\subsubsection{Collision of two strong shocks}

\begin{figure}[tbp]
{%
\centerline{\epsfig{file=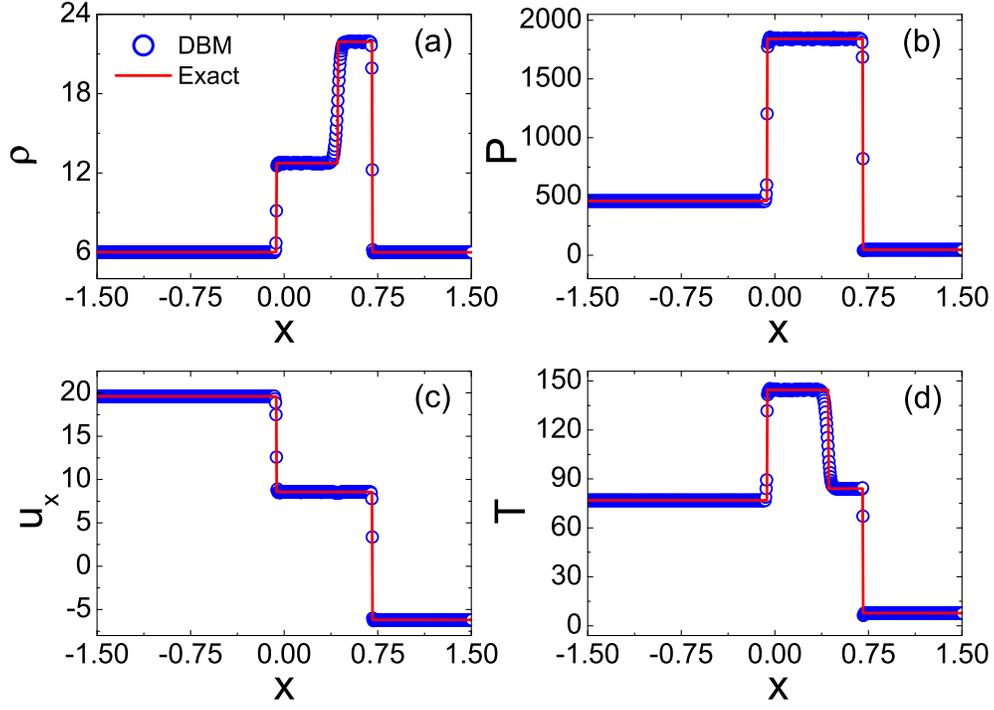,bbllx=7pt,bblly=3pt,bburx=576pt,bbury=434pt,
width=0.8\textwidth,clip=}}}
\caption{Comparisons between DBM simulations and the exact solutions for the
collision of two strong shocks, where $t=0.05$ and $\protect\gamma=1.67$.
(a) Density, (b)
pressure, (c) velocity, and (d) temperature.} 
\end{figure}

To further examine the robustness, precision, and adaptability of the model
for compressible flow with strong shock strength, we consider the collision
of two strong shocks described by
\begin{equation}
\left\{
\begin{array}{c}
(\rho ,T,u_{x},u_{y})|_{L}=(5.99924,76.8254,19.5975,0.0), \\
(\rho ,T,u_{x},u_{y})|_{R}=(5.99242,7.69222,-6.19633,0.0).%
\end{array}%
\right.
\end{equation}
With respect to the former two tests, this is generally regarded as a more
challenging one. Analytical solution contains a contact discontinuity moving
rightward, a right-shock spreading to the right side, and a left-shock
propagating rightward very slowly creating additional difficulties to the
numerical scheme. Figure 5 displays comparisons between DBM results and the
corresponding exact solutions, where $t=0.05$, $\gamma=1.67$. Parameters
used here are $\Delta x=\Delta y= 4 \times 10^{-3}$, $\Delta t=10^{-4}$, $%
\tau=5\times 10^{-5}$, $c=9$, and $\eta_{0}=30$. One can see that our
results are in satisfying agreement with the theoretical solutions with very
correct propagation of the shocks. Successful simulation of this aggressive
test manifests that the proposed model is robust, accurate and applicable to
compressible flows with strong shock wave interaction.

\subsection{Performance of the DBM for describing higher-order TNE effects}

To evaluate whether the model can describe TNE effects at various levels and
whether the model can reproduce accurate viscous stress and heat flux for
compressible flows over a wide range of Knudsen numbers and Mach numbers, a
series of simulations for head-on collision between two shocks have been
conducted. The initial configurations are
\begin{equation}
\rho (x,y)=\frac{{\rho_{L}+\rho_{R}}}{2}-\frac{{\rho_{L}-\rho_{R}}}{2}\tanh (%
\frac{x-N_{x} \Delta x/2}{L{_{\rho }}})\text{,}
\end{equation}%
\begin{equation}
u_{x}(x,y)=-{u_{0}}\tanh (\frac{x-N_{x} \Delta x/2}{L{_{u}}})\text{,}
\end{equation}%
where $L_{\rho }$ and $L_{u}$ are the widths of density and velocity
transition layers, respectively. ${\rho _{L}}$ (${\rho _{R}}$) is the
density away from the interface of the left (right) fluid. The whole
computational domain is a rectangle with length $1.5$ and height $0.006$,
divided into $1000\times 4$ uniform meshes.

\subsubsection{ Viscous stress}

According to the analytical expressions of TNE manifestations, two factors
control their strengths and structures, the relaxation time $\tau$ and the
gradient force induced by gradients of macroscopic quantities. In the first
three sets of simulations, we fix variables as ${\rho _{L}}=2{\rho _{R}}=2$,
$P_L=P_R=2$, ${u}_{y}=0$, $L_{\rho }=L_{u}=20$, then adjust $\tau $ and $%
u_{0}$, resulting in three types of viscous stresses. Figure 6 depicts the
details at $t=0.025$, where two DBMs are used: the D2V16 model at the NS
level [left column, satisfies the former 7 kinetic moments, Eqs. (\ref{M0})-(%
\ref{M31}) and Eqs. (\ref{M3})-(\ref{M42})], and the D2V26 model at the
Burnett level [right column, satisfies all needed kinetic moments, Eqs. (\ref%
{M0})-(\ref{M31}) and Eqs. (\ref{M3})-(\ref{M53})]. For comparisons, the
analytical solutions with the first and second order accuracies calculated
from Eqs. (\ref{vis1}), (\ref{heat1}), (\ref{A1}), and (\ref{A4}) are
plotted in each panel by dashed and solid lines, respectively.

\begin{figure}[tbp]
{%
\centerline{\epsfig{file=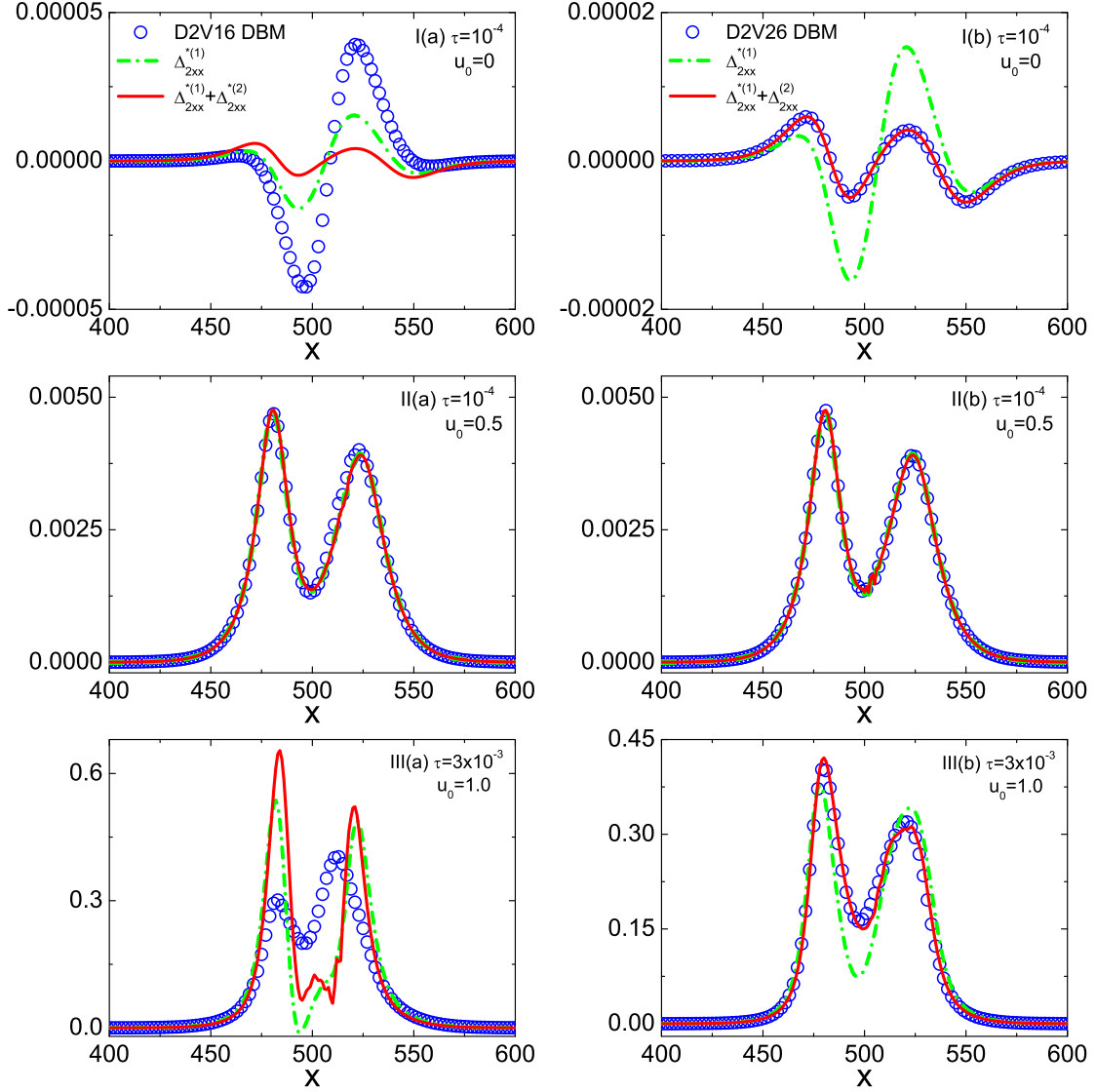,bbllx=108pt,bblly=218pt,bburx=567pt,bbury=683pt,
width=0.9\textwidth,clip=}}}
\caption{ (Color online) Viscous stress calculated from D2V16 (left column)
and D2V26 (right column) DB simulations (scatters) for the weak (I),
moderate (II), and strong (III) cases, where dashed and solid lines indicate
analytical solutions with the first and second order accuracies,
respectively.}
\end{figure}

\begin{figure}[tbp]
{%
\centerline{\epsfig{file=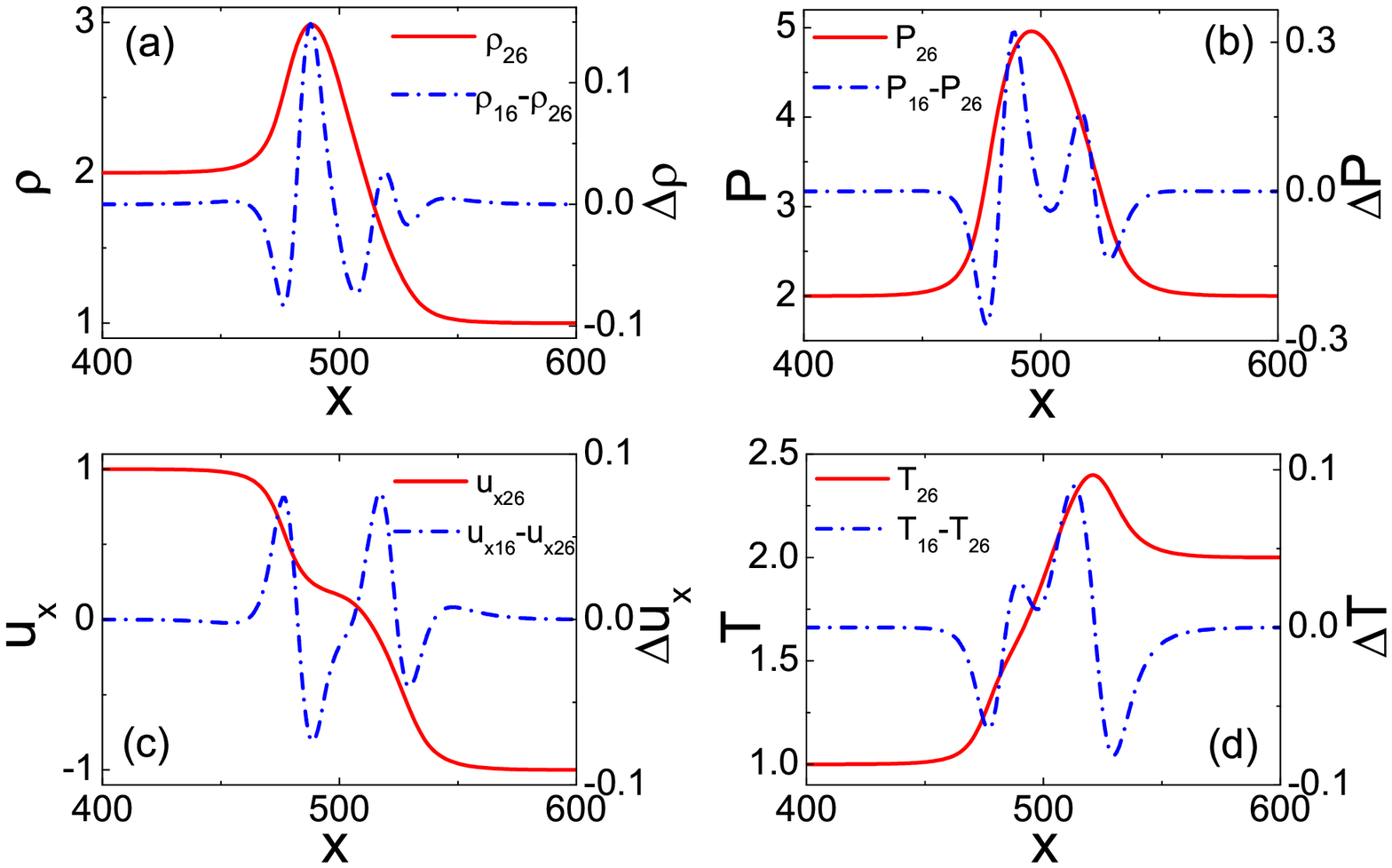,bbllx=7pt,bblly=0pt,bburx=574pt,bbury=358pt,
width=0.8\textwidth,clip=}}}
\caption{ (Color online) Hydrodynamic quantities calculated from the D2V26
model, and the corresponding differences between the D2V26 and D2V16 models
at $t=0.025$ for case III. (a) Density, (b) pressure, (c)
velocity, and (d) temperature.} 
\end{figure}

Figure 6 qualitatively reveals the common features during the procedure
deviating from thermodynamic equilibrium: (i) TNE effects are mainly around
the contact interface where the gradients of macroscopic quantities are
pronounced and exactly attain their local maxima (minima) at the points of
the maxima ($\bm{\nabla}\rho, \bm{\nabla}T, \bm{\nabla}u_{x})_{\max }$, for
example at $x=477$ and $522$; while they tend to vanish where the TNE
driving force is nearly zero. Behaviors of TNE can be well interpreted by
our theoretical formulations. (ii) For all cases, the first-order NOMF $%
\Delta _{2xx}^{\ast (1)}$, linearly proportional to $\tau $, is larger than
the second-order NOMF $\Delta _{2xx}^{\ast(2)}$, proportional to $\tau^{2}$,
numerically manifesting that $\Delta _{2xx}^{\ast (1)}$ is the leading part
of $\Delta_{2xx}^{\ast }$ and the appropriateness of NS model as a
coarse-grained model for compressible flows.

Apart from similarities, the following distinctive differences between
various cases and models deserve more attention. Different relaxation times
and shock intensities generate different TNE amplitudes. For case I (first row), due
to lack of velocity gradient ($\mathbf{u}=0$), at the beginning, viscous
stress is only induced by gradients of density and temperature. Therefore, $%
\Delta_{2xx}^{\ast (2)}>\Delta _{2xx}^{\ast (1)}\simeq 0$. Afterwards, the
density and temperature gradients stimulate velocity gradients, then $%
\Delta_{2xx}^{\ast (1)}$ becomes gradually larger than $\Delta _{2xx}^{\ast
(2)}$. In both cases, smaller $\tau$ and the initially static fluid lead to
weaker TNE effects. Nevertheless, even for such a very tiny TNE amplitude,
remarkable discrepancies appear between the D2V16 simulations and the
theoretical predictions, regardless of the first-order or the second-order
one [(see panel I(a)].
On the contrary, the D2V26 result agrees well with the theoretical
solution at the second-order $\Delta_{2xx}^{\ast(1)}+\Delta_{2xx}^{\ast(2)}$ [(see panel I(b)].
The D2V16 model is accurate at the NS level, without considering the
second-order TNE effects, and therefore it is not not suitable for
simulating cases when $\Delta_{2xx}^{\ast (2)}$ is as important as $\Delta
_{2xx}^{\ast (1)}$. For case II (second row),
we increase the intensity of TNE through increasing the collision velocity.
As a result, $\Delta_{2xx}^{\ast }$ is $100$ times larger than that in case (I), and $\Delta_{2xx}^{\ast (2)}$ is
negligible compared with $\Delta_{2xx}^{\ast (1)}$, demonstrating that the
velocity gradient acts as the dominating factor for TNE intensity. Excellent
agreements between DBM simulations and theoretical solutions are found for
both models [see panels II(a) and II(b)].
Further increase in relaxation time and collision velocity give
rise to more prominent TNE phenomena and more remarkable deviation from the
Maxwellian distribution, as shown in case III (third row). We observe that, larger
velocity not only induces a huge first-order TNE $\Delta_{2xx}^{\ast (1)}$,
but also prominently triggers the gradients in density and temperature (see
Fig. 7 for more details), and consequently, results in considerable
second-order TNE $\Delta_{2xx}^{\ast(2)}$. The D2V16 model fails to tame
such strong TNE behaviors, while the D2V26 model succeeds [see panels III(a) and III(b)].

\begin{figure}[tbp]
{%
\centerline{\epsfig{file=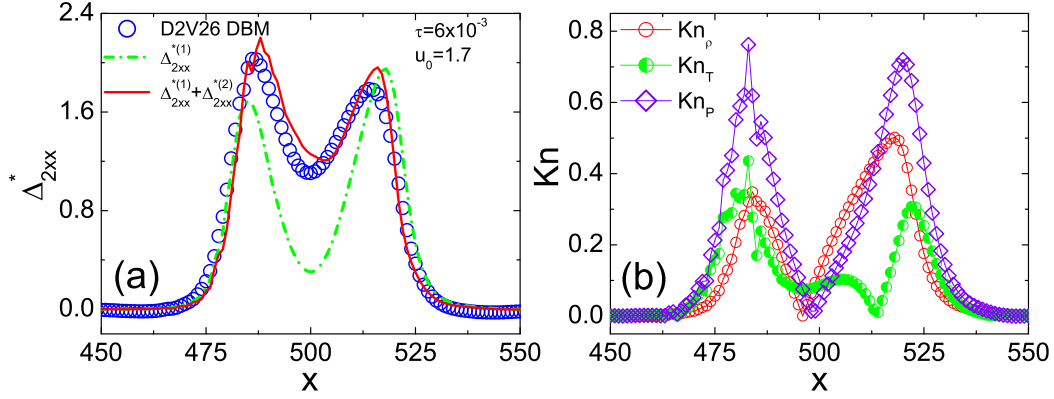,bbllx=2pt,bblly=2pt,bburx=580pt,bbury=221pt,
width=0.85\textwidth,clip=}}}
\caption{ (Color online) Viscous stress for the very strong case (a) and the
local Knudsen numbers calculated from pressure, density and temperature (b).}
\end{figure}

\begin{figure}[tbp]
{%
\centerline{\epsfig{file=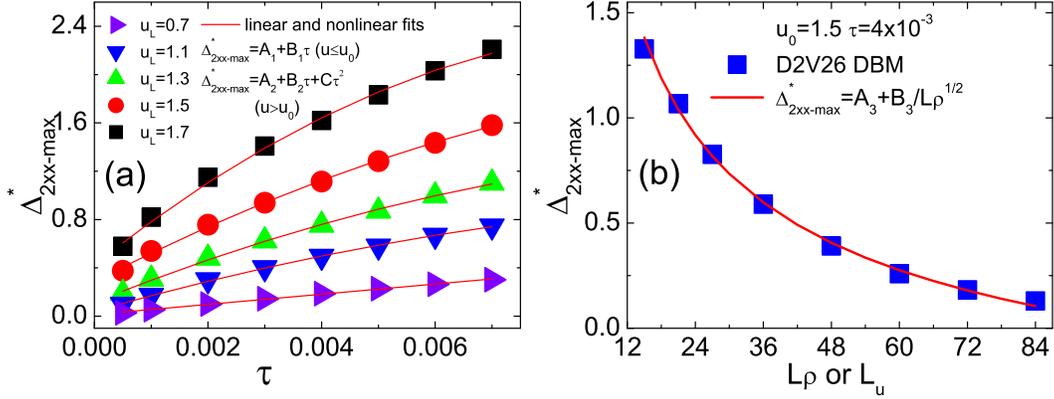,bbllx=3pt,bblly=2pt,bburx=580pt,bbury=230pt,
width=0.85\textwidth,clip=}}}
\caption{ (Color online) Effects of shock intensity (a) and interface width 
(b) on TNE effects.}
\end{figure}

Usually, the local Knudsen number, defined as the ratio of molecular
mean-free-path to a local characteristic length scale $Kn=\lambda /L$, is
one of the main parameters employed to describe the level of
non-equilibrium, where $\lambda =c_{s}\tau $, $c_{s}$ is the local speed of
sound, $L$ can be defined in terms of the macroscopic gradients, e.g., $%
L=\phi /|\bm{\nabla}\phi |$. The maxima $Kn_{\max}$ for cases I, II, and III are $%
0.0018$, $0.003$, and $0.15$, respectively, all beyond the application scope of
the NS model. Actually, the D2V26 model has been extended into the early
transition regime. It is also interesting to note that, the D2V16 model is
more reliable and more powerful to study case II than case I. Thus, from
this point of view, Knudsen number is not sufficient enough to describe the
TNE extent for cases with small Mach numbers. To complement this deficiency,
we introduce another dimensionless parameter to characterize the relative
TNE intensity, $R_{\text{TNE}}=|\Delta _{2}^{\ast (2)}/\Delta _{2}^{\ast
(1)}|$. For the three cases, $R_{\text{TNE}}=0.69$, $0.01$, $0.42$, respectively.
Consequently, higher-order DBMs are needed for cases I and III, even
though the TNE intensity is weak in case I. It is convenient to generalize
the definition as $R_{\text{TNE}}=|\Delta _{m,n}^{\ast (N+1)}/\Delta
_{m,n}^{\ast (N)}|$, where $\Delta_{m,n}^{\ast (N+1)}$ ($\Delta
_{m,n}^{\ast (N)}$) is the ($N+1$)-th ($N$-th) order TNE. Meanwhile, we can
define the TNE discrepancy between DBM simulation and the corresponding
theoretical analysis, $\varrho=\Delta_{\text{DBM}}-\Delta _{\text{Exact}}$.
These two measures provide as effective physical criteria to assess
whether the current DBM is appropriate or not. In real simulations, only
when the $R_{\text{TNE}}$ and/or $\varrho$ is small enough, the current DBM
is suitable for describing the current problem; otherwise, higher-order TNE
effects should be taken into account in the modeling and higher-order DBM
should be constructed.

We also stress that the exact calculation of viscous stress and heat flux
are of great importance for simulating high-speed, non-equilibrium
compressible flows, because the transport and dissipation of kinetic energy
and momentum resulting in complex mesoscopic structures (such as shock wave
interface, material interface, Mach stem, etc.) depend strongly on them.
More importantly, accurate viscous stress and heat flux are required in
order to obtain accurate hydrodynamic quantities, as demonstrated by Fig. 7,
where $\phi_{16}-\phi_{26}$ indicates hydrodynamic quantities differences
between the D2V16 and D2V26 models for case III. It is clear that, the
differences, up to $10\%$ of the exact solutions, are around the highly
non-equilibrium regimes. The inaccuracies of the D2V16 model are due to the
lack of some necessary kinetic moments required for recovering $f^{(2)}$.

To further examine the reliability of D2V26 model in describing much
stronger TNE effects, we increase $\tau$ to $6\times 10^{-3}$ and $u_{0}=1.7$
. Shown in Fig. 8 are viscous stress [panel (a)] for the very accentuated case and the
local Knudsen numbers [panel (b)] calculated from pressure, density and temperature,
respectively. Good agreement between the DBM simulation and the second-order
theoretical solution can be found. The maximum Knudsen number calculated
from density exceeds $0.5$, and the one calculated from pressure is as high
as $0.8$. When the strength of TNE further increases, the presented model
loses its effectiveness and effects of $f^{(3)}$ should be taken into
account.

\begin{figure}[tbp]
{%
\centerline{\epsfig{file=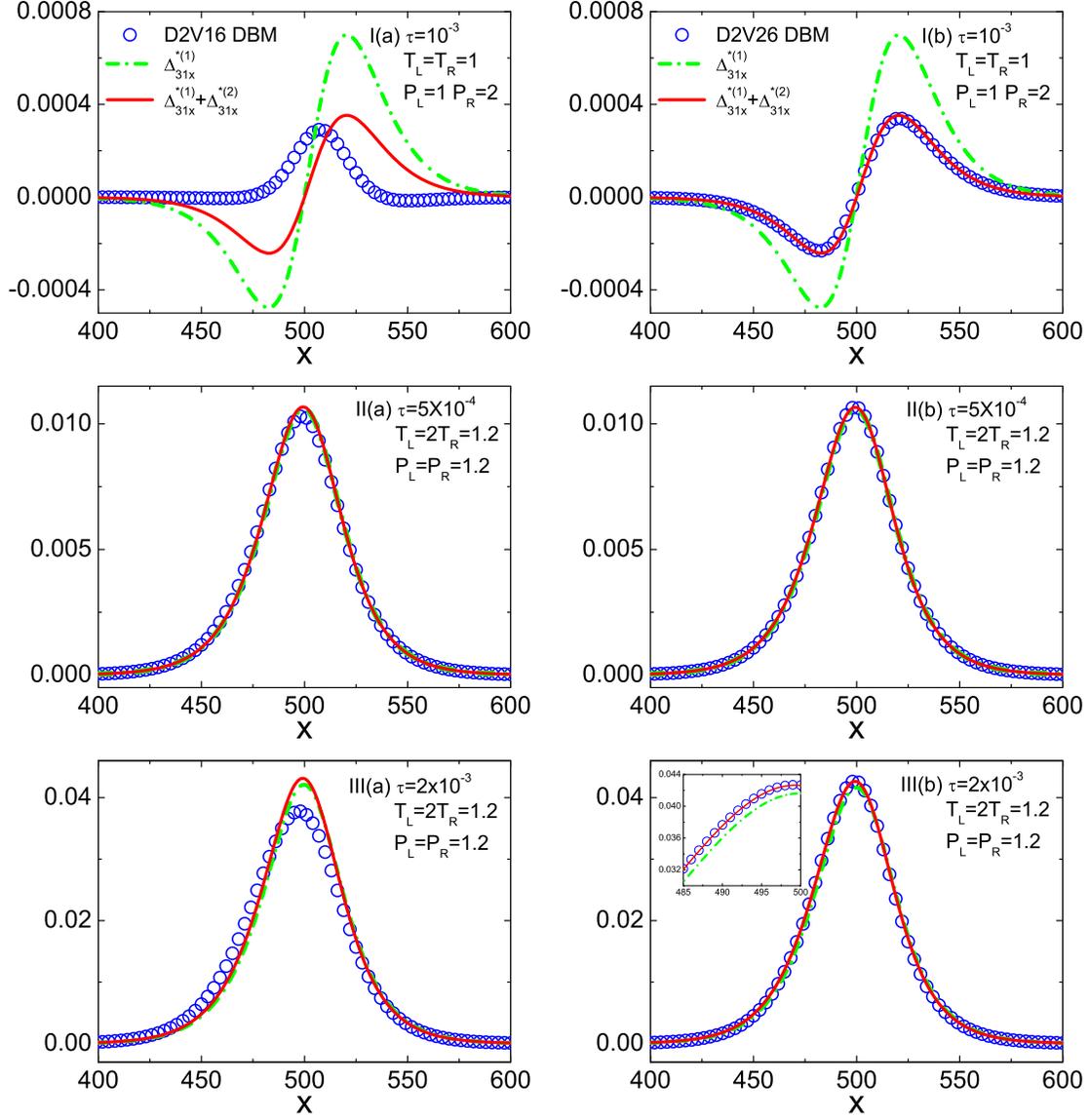,bbllx=111pt,bblly=211pt,bburx=570pt,bbury=690pt,
width=0.9\textwidth,clip=}}}
\caption{ (Color online) Heat flux calculated from D2V16 (left column) and
D2V26 (right column) DBM simulations for the weak (I), moderate (II) and
strong (III) cases, where dashed and solid lines indicate analytical
solutions with the first and second order accuracies, respectively.}
\end{figure}

Effects of shock intensity and interface width on TNE manifestations are
investigated similarly. As plotted in Fig. 9(a), the maximum non-equilibrium
stress increases with both $\tau$ and $u_{0}$. The relationship between $%
\Delta _{2xx-{\max }}^{\ast }$ and $\tau$ can be further divided into two
cases: linear and nonlinear. When $u_{0}$ is less than a critical value $%
u_{c}$, say $0.7$, $\Delta _{2xx{-\max }}^{\ast }$ increases linearly with $%
\tau $, $\Delta _{2xx{-\max }}^{\ast }=A_{1}+B_{1}\tau $; when $u_{0}>u_{c}$%
, a nonlinear fitting is more approximate, $\Delta _{2xx-{\max }}^{\ast
}=A_{2}+B_{2}\tau +C_{2}\tau ^{2}$, demonstrating the necessity of a
higher-order constitutive relations for cases far-away-from-equilibrium.
Conversely, the interface width effects decrease the maximum of $%
\Delta_{2xx}^{\ast }$ approximately in the following way, $\Delta _{2xx{%
-\max }}^{\ast }=A_{3}+\frac{B_{3}}{L\bigskip ^{1/2}}$, with $A_{3}=-0.81$
and $B_{3}=8.47$, as shown in Fig. 9(b). This conclusion is consistent with
the effects of surface tension that controls the width of hydrodynamic
quantities in multiphase flows \cite{2015-AG-SM}.
Physically, the interface
width lowers the gradient force and suppresses the TNE intensity.

\subsubsection{Heat flux}

The viability of the D2V26 model for describing higher-order heat flux is
verified in a similar way. Consistently, three cases are considered, with
the following initial variables, case I: $T_{L}=T_{R}=1$, $P_{L}=1$, $%
P_{R} $=2, $\tau =10^{-3}$; case II: $T_{L}=2T_{R}=1.2$, $P_{L}=P_{R}=1.2$%
, $\tau =5\times {10^{-4}}$; case III: $T_{L}=2T_{R}=1.2$, $%
P_{L}=P_{R}=1.2 $, $\tau =2\times {10^{-3}}$. Collision velocity for all
cases is fixed to be $u_{0}=0.5$. Figure 10 presents the details, where $%
t=10^{-3}$ in case II and $t=9\times 10^{-3}$ in the other two cases. For
the first case, temperature is initially homogeneous, thus $\Delta
_{3,1x}^{\ast (1)}$ approaches nearly zero at the beginning stage. The
second-order heat flux $\Delta _{3,1x}^{\ast (2)}$ is motivated exclusively
by a pressure difference. After that, gradients appear in each quantity
resulting in the emergence of the first-order heat flux. At the moment shown
in case I, the relative intensity $R_{\text{TNE}}=\Delta _{3,1x}^{\ast
(2)}/\Delta _{3,1x}^{\ast (1)}$ is about $0.98$. As excepted, D2V16 model
fails to predict this situation although with weak TNE intensity [see panel I(a)]. Through
enlarging gradient in temperature in case II, $\Delta _{3,1x}^{\ast}$ is
overwhelmed by $\Delta _{3,1x}^{\ast (1)}$, as reported in the second row of Fig. 6. For
this case, the two models recover favorable results [see panels II(a) and II(b)]. The deficiency of the
D2V16 model and the sufficiency of the D2V26 model for portraying TNE with
higher amplitude, has been witnessed by case III [see panels III(a) and III(b)], again.

\section{Conclusions and remarks}

A framework for constructing the trans-scale DBM that aims to investigate
high-speed compressible flows ranging from continuum to transition regime,
is presented. In this framework, the specific forms of the extremely complex
Burnett, even super-Burnett, equations are not needed. To access
higher-order non-equilibrium effects, the extension of the framework and the
construction of corresponding DBM are more convenient and straightforward
than the extended hydrodynamic equations; the complexity of the DBM
increases only mildly, as opposed to the sharp raise of complexity of the
thermo-hydrodynamic equations. Through switching the effective parameter
that controls the TNE extent, one can perform multi-scale simulations over a
wide range of Knudsen number under the same framework without message
passing between models at different scales. As a model example, a
two-dimensional DBM with $26$ discrete velocities at Burnett level is
formulated, verified and validated. As by-products, the linear and
non-linear constitutive relations for the hydrodynamic modeling are derived,
which contribute to improve the macroscopic modeling. To better characterize
the non-equilibrium flows and understand the conditions under which the DBMs
at various levels must be used, besides some higher-order kinetic moments of
$(f -f^{(0)})$ and the Knudsen number, two additional criteria, i.e., (i)
the relative TNE strength, describing the relative strength of the $(N+1)$%
-th order TNE to the $N$-th order one, and (ii) the TNE discrepancy between
DBM simulation and corresponding theoretical analysis, are defined. Whether
or not the higher-order TNE effects should be taken into account in the
modeling process and which level of DBM should be utilized, depends on the
relative strength of the higher-order TNE with respect to the current order
and/or the TNE discrepancy, instead of the value of Knudsen number itself.

\section*{Acknowledgements}

The authors sincerely thank the anonymous reviewers for their valuable
comments and suggestions, which are very helpful for revising the
manuscript. Also, we warmly thank Dr. Chuandong Lin, Dr. Ge Zhang, Dr.
Huilin Lai, and Dr. Bohai Chen for many instructive discussions. We
acknowledge support from the National Natural Science Foundation of China
(11475028, 11772064 and 11602162), Science Challenge Project
(JCKY2016212A501), Natural Science Foundation of Hebei Province (A2017409014
and A201500111), Natural Science Foundations of Hebei Educational Commission
(ZD2017001) and FJKLMAA, Fujian Normal University.

\appendix

\section{Formulations of the second-order viscous stress and heat flux}
\begin{eqnarray}
\Delta_{2xx}^{\ast (2)} &=&2n_{2}^{-2}\tau ^{2}\{\rho RT[n_{-2}n_{1}\left(
\partial _{x}u_{x}\right) ^{2}+n_{1}n_{2}\left( \partial _{y}u_{x}\right)
^{2}-4n\,\partial _{x}u_{x}\partial _{y}u_{y}-n_{2}\left( \partial
_{x}u_{y}\right) ^{2}  \notag \\
&&-n_{-2}\,\left( \partial _{y}u_{y}\right) ^{2}]+\rho
R^{2}[n_{1}n_{2}\left( \partial _{x}T\right) ^{2}-n_{2}\left( \partial
_{y}T\right) ^{2}]-R^{2}T^{2}[n_{1}n_{2}\frac{\partial ^{2}}{\partial x^{2}}%
\rho -n_{2}\frac{\partial ^{2}}{\partial y^{2}}\rho ]  \notag \\
&&+\frac{R^{2}T^{2}}{\rho }[n_{1}n_{2}(\partial _{x}\rho
)^{2}-n_{2}(\partial _{y}\rho )^{2}]\},  \label{A1}
\end{eqnarray}%
\begin{eqnarray}
\Delta_{2xy}^{\ast (2)} &=&2\tau ^{2}[\,n_{2}^{-1}\rho T(n\partial
_{x}u_{x}\partial _{x}u_{y}+n\partial _{y}u_{x}\partial _{y}u_{y}-2\partial
_{x}u_{y}\partial _{y}u_{y}-2\partial _{x}u_{x}\partial _{y}u_{x})  \notag \\
&&+\rho R^{2}\partial _{x}T\partial _{y}T-R^{2}T^{2}\frac{\partial ^{2}}{%
\partial x\partial y}\rho +\frac{R^{2}T^{2}}{\rho }\partial _{x}\rho
\partial _{y}\rho ],  \label{A2}
\end{eqnarray}%
\begin{eqnarray}
\Delta_{2yy}^{\ast (2)} &=&-2n_{2}^{-2}\tau ^{2}\{\rho RT[n_{-2}\left(
\partial _{x}u_{x}\right) ^{2}+n_{2}\left( \partial _{y}u_{x}\right)
^{2}+4n\,\partial _{x}u_{x}\partial _{y}u_{y}-n_{1}n_{2}\left( \partial
_{x}u_{y}\right) ^{2}  \notag \\
&&-n_{-2}n_{1}\,\left( \partial _{y}u_{y}\right) ^{2}]+\rho
R^{2}[n_{2}\left( \partial _{x}T\right) ^{2}-n_{1}n_{2}\left( \partial
_{y}T\right) ^{2}]-R^{2}T^{2}[n_{2}\frac{\partial ^{2}}{\partial x^{2}}\rho
-n_{1}n_{2}\frac{\partial ^{2}}{\partial y^{2}}\rho ]  \notag \\
&&+\frac{R^{2}T^{2}}{\rho }[n_{2}(\partial _{x}\rho
)^{2}-n_{1}n_{2}(\partial _{y}\rho )^{2}]\},  \label{A3}
\end{eqnarray}%
\begin{eqnarray}
\Delta_{3,1x}^{\ast (2)} &=&n_{2}^{-1}\tau ^{2}\{\rho R^{2}T^{2}[n_{-2}\frac{%
\partial ^{2}}{\partial x^{2}}u_{x}+n_{2}\frac{\partial ^{2}}{\partial y^{2}}%
u_{x}-4\,\frac{\partial ^{2}}{\partial x\partial y}u_{y}]+\rho
R^{2}T[(n_{2}^{2}+4n)\partial _{x}u_{x}\partial _{x}T  \notag \\
&&+n_{2}n_{6}\partial _{y}u_{x}\partial _{y}T-2n_{6}\partial
_{y}u_{y}\partial _{x}T+2n_{2}\partial _{x}u_{y}\partial _{y}T\,]\},
\label{A4}
\end{eqnarray}%
\begin{eqnarray}
\Delta_{3,1y}^{\ast (2)} &=&n_{2}^{-1}\tau ^{2}\{\rho R^{2}T^{2}[n_{2}\frac{%
\partial ^{2}}{\partial x^{2}}u_{y}+n_{-2}\frac{\partial ^{2}}{\partial y^{2}%
}u_{y}-4\,\frac{\partial ^{2}}{\partial x\partial y}u_{x}]+\rho
R^{2}T[(n_{2}^{2}+4n)\partial _{y}u_{y}\partial _{y}T  \notag \\
&&+n_{2}n_{6}\partial _{x}u_{y}\partial _{x}T-2n_{6}\partial
_{x}u_{x}\partial _{y}T\,+2n_{2}\partial _{y}u_{x}\partial _{x}T]\},
\label{A5}
\end{eqnarray}%
where $n_{a}=n+a$.


\begin{thebibliography}{999}
\bibitem{Xu-SciChina} A. Xu, G. Zhang, Y. Ying, and C. Wang, Sci.
China-Phys. Mech. Astron. \textbf{59}, 650501 (2016).

\bibitem{Ju} Y. G. Ju, Adv. Mech. \textbf{44}, 201402 (2014).

\bibitem{Sterilization} G. R, Nigri, S. Tsai, S. Kossodo, P. Waterman, P.
Fungaloi, D. C. Hooper, A. G. Doukas, and G. M. Lamuraglia, Laser. Surg.
Med. \textbf{29}, 448 (2001).

\bibitem{medical-treatment} J. J. Rassweiler, T. Knoll, K. U. K\"{o}hrmann,
J. A. Mcateer, J. E. Lingeman, R. O. Cleveland, M. R. Bailey, and C.
Chaussy, Eur. Urol. \textbf{59}, 784 (2011).

\bibitem{food} N. Boussetta, E. Vorobiev, T. Reess, A. De Ferron, L.
Pecastaing, R. Ruscassi\'{e}, and J.-L. Lanoisell\'{e}, Innov. Food Sci.
Emerg. \textbf{16}, 129 (2012).

\bibitem{2004-Li-JCP} Z. H. Li and H. X. Zhang, J. Comput. Phys. \textbf{193}%
, 708 (2004).

\bibitem{2009-Li-JCP} Z. H. Li and H. X. Zhang, J. Comput. Phys. \textbf{228}%
, 1116 (2009).

\bibitem{2015-Li-PAS} Z. H. Li, A. P. Peng, H. X. Zhang, and J. Y. Yang,
Prog. Aerosp. Sci. \textbf{74}, 81 (2015).

\bibitem{2015-Wang} Z. H. Wang, \textit{Theoretical Modelling of Aeroheating
on Sharpened Noses Under Rarefied Gas Effects and Nonequilibrium Real Gas
Effects}, (Springer, New York, 2014).

\bibitem{He-SciChina} H. Liu, W. Kang, H. Duan, P. Zhang, and X. He, Sci.
China-Phys. Mech. Astron. \textbf{47}, 070003 (2017) (in Chinese).

\bibitem{WLF} L. Wang, W. Ye, X. He, J. Wu, Z. Fan, C. Xue, H. Guo, W. Miao,
Y. Yuan, J. Dong, G. Jia, J. Zhang, Y. Li, J. Liu, M. Wang, Y. Ding, and W.
Zhang, Sci. China-Phys. Mech. Astron. \textbf{60}, 055201 (2017).

\bibitem{2004-JFM} R. Balakrishnan, J. Fluid Mech. \textbf{503}, 201 (2004).

\bibitem{Succi-Science} H. Chen, S. Kandasamy, S. Orszag, R. Shock, S.
Succi, and V. Yakhot, Science \textbf{301}, 633 (2003).

\bibitem{Caflisch1995} W. J. Morokoff and R. E. Caflisch, J. Comput. Phys.
\textbf{122}, 218 (1995).

\bibitem{Caflisch1998} R. E. Caflisch, Acta Numer. \textbf{7}, 1 (1998).

\bibitem{Caflisch1999} L. Pareschi and R. E. Caflisch, J. Comput. Phys.
\textbf{154}, 90 (1999).

\bibitem{Succi-Book} S. Succi, \textit{The Lattice Boltzmann Equation for
Fluid Dynamics and Beyond}, (Oxford University Press, New York, 2001).

\bibitem{Succi-2038} S. Succi, EPL \textbf{109}, 50001 (2015).

\bibitem{Struchtrup} H. Struchtrup, \textit{Macroscopic Transport Equations
for Rarefied Gas Flows}, (Springer, New York, 2005).

\bibitem{XuKun} K. Xu, \textit{Direct Modeling for Computational Fluid
Dynamics: Construction and Application of Unified Gas-kinetics Schemes},
(World Scientific Publishing, Beijing, 2015).

\bibitem{Guo-Shu} Z. Guo and C. Shu, \textit{Lattice Boltzmann Method and
its Applications in Engineering}, (World Scientific Publishing, Beijing,
2013).

\bibitem{2014-Zhang-JFM} L. Wu, J. M. Reese, and Y. Zhang, J. Fluid Mech.
\textbf{746}, 53 (2014).

\bibitem{2015-Zhang-JCP} L. Wu, J. Zhang, J. M. Reese, and Y. Zhang, J.
Comput. Phys. \textbf{298}, 602 (2015).


\bibitem{2016-Shu-JCP} L. M. Yang, C. Shu, J. Wu, Y. Wang, J. Comput. Phys.
\textbf{306}, 291 (2016).


\bibitem{2013-Shu-JCP} L. M. Yang, C. Shu, J. Wu, N. Zhao, and Z. L. Lu, J.
Comput. Phys. \textbf{255}, 540 (2013).

\bibitem{2014-Shu-JCP} L. M. Yang, C. Shu, and J. Wu, J. Comput. Phys.
\textbf{274}, 611 (2014).

\bibitem{2016-Shu-PRE} L. M. Yang, C. Shu, and Y. Wang, Phys. Rev. E \textbf{%
93}, 033311 (2016).

\bibitem{2016-Shu-JCP2} L. M. Yang, C. Shu, Y. Wang, and Y. Sun, J. Comput.
Phys. \textbf{319}, 129 (2016).

\bibitem{2015-Guo-PRE} Z. Guo, R. Wang, and K. Xu, Phys. Rev. E, \textbf{91}%
, 033313 (2015).

\bibitem{2016-Guo-PRE} P. Wang, L.-P. Wang and Z. Guo, Phys. Rev. E \textbf{%
94}, 043304 (2016).

\bibitem{2016-Xu-JCP} C. Liu, K. Xu, Q. Sun, and Q. Cai, J. Comput. Phys.
\textbf{314}, 305 (2016).

\bibitem{2017-Xu-JCP} T. Xiao, Q. Cai, and K. Xu, J. Comput. Phys. \textbf{%
332}, 475 (2017).


\bibitem{2003-Struchtrup-POF} H. Struchtrup and M. Torrilhon, Phys. Fluids
\textbf{15}, 2668 (2003).

\bibitem{2004-Struchtrup-JFM} M. Torrilhon and H. Struchtrup, J. Fluid Mech.
\textbf{513}, 171 (2004).

\bibitem{2007-PRL-Struchtrup} H. Struchtrup and M. Torrilhon, Phys. Rev.
Lett. \textbf{99}, 014502 (2007).

\bibitem{2017-Struchtrup-POF} M. Yu. Timokhin, H. Struchtrup, A. A.
Kokhanchik, and Ye. A. Bondar, Phys. Fluids \textbf{29}, 037105 (2017).

\bibitem{2009-Gu-JFM} X. Gu, and D. R. Emerson, J. Fluid Mech. \textbf{636},
177 (2009).

\bibitem{2008-Fox-JCP} R. O. Fox, J. Comput. Phys. \textbf{227}, 6313 (2008).

\bibitem{2009-Fox-JCP} R.O. Fox, J. Comput. Phys. \textbf{228}, 7771 (2009).

\bibitem{2012-Fox-JCP} M. Icardi, P. Asinari, D. L. Marchisio, S. Izquierdo,
and R. O. Fox, J. Comput. Phys. \textbf{231 }, 7431 (2012).


\bibitem{1992-BSV} R. Benzia, S. Succi and M. Vergassolac, Phys. Rep.
\textbf{222}, 145 (1992).

\bibitem{2001-Succi} S. Succi, O. Filippova, G. Smith, and E. Kaxiras,
Comput. Sci. Eng., \textbf{3}, 26 (2001).

\bibitem{2002-Succi} S. Succi, I. V. Karlin and H. Chen, Rev. Mod. Phys.
\textbf{74}, 1203 (2002).


\bibitem{1998-Wagner-PRL} A. J. Wagner and J. M. Yeomans, Phys. Rev. Lett.
\textbf{80}, 1429 (1998).

\bibitem{1999-Wagner-PRE} A. J. Wagner and J. M. Yeomans, Phys. Rev. E
\textbf{59}, 4366 (1999).

\bibitem{2007-Yeomans-PRE} D. Marenduzzo, E. Orlandini, M. E. Cates, and J.
M. Yeomans, Phys. Rev. E \textbf{76}, 031921 (2007).

\bibitem{2016-Yeomans-PRL} A. Doostmohammadi, S. P. Thampi, and J. M.
Yeomans, Phys. Rev. Lett. \textbf{117}, 048102 (2016).

\bibitem{2016-Wagner-PRE} A. J. Wagner and K. Strand, Phys. Rev. E \textbf{94%
}, 033302 (2016).


\bibitem{2005-Ansumali-PRL} S. Ansumali and I. V. Karlin, Phys. Rev. Lett.
\textbf{95}, 260605 (2005).

\bibitem{2017-Ansumali-PRL} M. Atif, P. K. Kolluru, C. Thantanapally, and S.
Ansumali, Phys. Rev. Lett. \textbf{119}, 240602 (2017).

\bibitem{2017-Huang-JFM} H. Huang and X.-Y. Lu, J. Fluid Mech. \textbf{822},
664 (2017).

\bibitem{2013-AG-EPL} Y. Gan, A. Xu, G. Zhang, and Y. Yang, EPL \textbf{103}%
, 24003 (2013).

\bibitem{2014-AG-PRE} C. Lin, A. Xu, G. Zhang, Y. Li, and S. Succi, Phys.
Rev. E \textbf{89}, 013307 (2014).

\bibitem{2015-AG-PRE} A. Xu, C. Lin, G. Zhang, and Y. Li, Phys. Rev. E
\textbf{91}, 043306 (2015).


\bibitem{2013-Shu-CF} L. M. Yang, C. Shu, and J. Wu, Comput. Fluids \textbf{%
79}, 190 (2013).

\bibitem{2016-Shu-CMA} L. M. Yang, C. Shuc, and J. Wu, Comput. Math. Appl.
\textbf{71}, 2069 (2016).

\bibitem{LiQing} Q. Li, K. H. Luo, Q. J. Kang, Y. L. He, Q. Chen, and Q.
Liu, Prog. Energy Combust. Sci. \textbf{52}, 62 (2016).

\bibitem{2013-JSM} B. I. Green and P. Vedula, J. Stat. Mech: Theory Exp.
\textbf{2013}, P07016 (2013).


\bibitem{Succi-DBM} M. La Rocca, A. Montessori, P. Prestininzi, and S.
Succi, J. Comput. Phys. \textbf{284}, 117 (2015).

\bibitem{2015-AG-SM} Y. Gan, A. Xu, G. Zhang, and S. Succi, Soft Matter
\textbf{11}, 5336 (2015).

\bibitem{2016-AG-PRE} H. Lai, A. Xu, G. Zhang, Y. Gan, Y. Ying, and S.
Succi, Phys. Rev. E \textbf{94}, 023106 (2016).

\bibitem{2016-AG-CF} C. Lin, A. Xu, G. Zhang, and Y. Li, Combust. Flame
\textbf{164}, 137 (2016).

\bibitem{2016-AG-CF2} Y. Zhang, A. Xu, G. Zhang, C. Zhu, and C. Lin,
Combust. Flame \textbf{173}, 483 (2016).

\bibitem{2017-Lin-PRE} C. Lin, A. Xu, G. Zhang, K. H. Luo, and Y. Li, Phys.
Rev. E \textbf{96}, 053305 (2017).

\bibitem{2017-Lin-SciRep} C. Lin, K. H. Luo, L. Fei, and S. Succi, Sci. Rep.
\textbf{7}, 14580 (2017).

\bibitem{2018-Lin-CF} C. Lin and K. H. Luo, Comput. Fluids \textbf{166}, 176
(2018).


\bibitem{2007-Ansumali-PRL} S. Ansumali, I. V. Karlin, S. Arcidiacono, A.
Abbas, and N. I. Prasianakis, Phys. Rev. Lett. \textbf{98}, 124502 (2007).

\bibitem{2008-Ansumali-PRE} W. P. Yudistiawan, S. Ansumali, and I. V.
Karlin, Phys. Rev. E \textbf{78}, 016705 (2008).

\bibitem{2017-Karlin-JFM} B. Dorschner, S. S. Chikatamarla, and I. V.
Karlin, J. Fluid Mech. \textbf{824}, 388 (2017).


\bibitem{2005-Succi-EPL} F. Toschi and S. Succi, Europhys. Lett. \textbf{69}%
, 549 (2005).

\bibitem{2005-Succi-POF} M. Sbragaglia and S. Succi, Phys. Fluids \textbf{17}%
, 093602 (2005).

\bibitem{2006-Succi-EPL} M. Sbragaglia and S. Succi, Europhys. Lett. \textbf{%
73}, 370 (2006).

\bibitem{2015-Succi-PRE} A. Montessori, P. Prestininzi, M. La Rocca, and S.
Succi, Phys. Rev. E \textbf{92}, 043308 (2015).

\bibitem{2016-Succi-JSC} G. Di Staso, H. J. H. Clercx, S. Succi, and F.
Toschi, J. Comput. Sci. \textbf{17}, 357 (2016).

\bibitem{2016-Succi-PT} G. Di Staso, H. J. H. Clercx, S. Succi, and F.
Toschi, Phil. Trans. R. Soc. A \textbf{374}, 20160226 (2016).


\bibitem{2006-Shan-JFM} X. Shan, X. F. Yuan, and H. Chen, J. Fluid Mech.
\textbf{550}, 413 (2006).

\bibitem{2011-Shan-PRE} J. Meng, Y. Zhang, and X. Shan, Phys. Rev. E \textbf{%
83}, 046701 (2011).

\bibitem{2013-Zhang-JFM} J. Meng, Y. Zhang, N. G. Hadjiconstantinou, G. A.
Radtke, and X. Shan, J. Fluid Mech. \textbf{718}, 347 (2013).


\bibitem{2005-Sofonea-PRE} V. Sofonea and R. F. Sekerka, Phys. Rev. E
\textbf{71}, 066709 (2005).

\bibitem{2005-Sofonea-JCP} V. Sofonea and R. F. Sekerka, J. Comput. Phys.
\textbf{207}, 639 (2005).

\bibitem{2009-Watari-PRE} M. Watari, Phys. Rev. E \textbf{79}, 066706 (2009).

\bibitem{2014-JCP-Zhang} J. Meng and Y. Zhang, J. Comput. Phys. \textbf{258}%
, 601 (2014).

\bibitem{2006-Zhang-PRE} Y. H. Zhang, X. J. Gu, R. W. Barber, and D. R.
Emerson, Phys. Rev. E \textbf{74}, 046704 (2006).

\bibitem{2008-Zhang-PRE} G. H. Tang, Y. H. Zhang, and D. R. Emerson, Phys.
Rev. E \textbf{77}, 046701 (2008).

\bibitem{He-FOP} H. Liu, W. Kang, Q. Zhang, Y. Zhang, H. Duan, and X. T. He,
Front. Phys. \textbf{11}, 115206 (2016).

\bibitem{He-PRE} H. Liu, Y. Zhang, W. Kang, P. Zhang, H. Duan, and X. T. He,
Phys. Rev. E \textbf{95}, 023201 (2017).

\bibitem{BGK} P. L. Bhatnagar, E. P. Gross, and M. Krook, Phys. Rev. \textbf{%
94}, 511 (1954).

\bibitem{ES} L. H. Holway, Phys. Fluids \textbf{9}, 1658 (1966).

\bibitem{Shakhov} E. M. Shakhov, Fluid Dyn. \textbf{3}, 95 (1972).

\bibitem{Rykov} V. A. Rykov, Fluid Dyn. \textbf{10}, 959 (1976).

\bibitem{Liu1990} G. Liu, Phys. Fluids A \textbf{2}, 277 (1990).

\bibitem{2012-Gan-CTP} Y. Gan, A. Xu, G. Zhang, and Y. Li, Commun. Theor.
Phys. \textbf{56}, 490 (2011).

\bibitem{IMEX} U. M. Ascher, S. J. Ruuth, and R. J. Spiteri, Appl. Numer.
Math. \textbf{25}, 151 (1997).

\bibitem{2011-Gan-PRE} Y. Gan, A. Xu, G. Zhang, and Y. Li, Phys. Rev. E
\textbf{83}, 056704 (2011).

\bibitem{QuKun} K. Qu, C. Shu, and Y. T. Chew, Phys. Rev. E \textbf{75},
036706 (2007).

\bibitem{HYL} Q. Li, Y. L. He, Y. Wang, and W. Q. Tao, Phys. Rev. E \textbf{%
76}, 056705 (2007).
\end{thebibliography}
\end{document}